\documentclass[12pt]{article}

\usepackage[dvips]{graphicx}
\usepackage{wrapfig}
\usepackage{amssymb}
\usepackage{amsmath}

\def\IZ{\mathbb {Z}}

\def\IR{\mathbb {R}}
\def\IC{\mathbb {C}}
\newcommand{\IP}{{\relax{\rm I\kern-.18em P}}}
\newcommand{\IF}{{\relax{\rm I\kern-.18em F}}}

\renewcommand{\thefootnote}{\fnsymbol{footnote}}

\makeatletter
 \renewcommand{\theequation}{%
       \thesection.\arabic{equation}}
 \@addtoreset{equation}{section}
\makeatother

\makeatletter
\def\eqnarray{%
 \stepcounter{equation}%
 \let\@currentlabel=\theequation
 \global\@eqnswtrue
 \global\@eqcnt\z@
 \tabskip\@centering
 \let\\=\@eqncr
 $$\halign to \displaywidth\bgroup\@eqnsel\hskip\@centering
 $\displaystyle\tabskip\z@{##}$&\global\@eqcnt\@ne
 \hfil$\displaystyle{{}##{}}$\hfil
 &\global\@eqcnt\tw@$\displaystyle\tabskip\z@{##}$\hfil
 \tabskip\@centering&\llap{##}\tabskip\z@\cr}
\makeatother

\begin{document}
%
\begin{titlepage}
\begin{center}
\vspace*{1cm}
{\Large \bf
Deformed planar topological open string amplitudes\\[3mm]
on Seiberg-Witten curve}
\vskip 1.5cm
{\large Masahide Manabe$^{1,2}$\footnote[1]{e-mail: masahidemanabe@gmail.com}}
\vskip 1.0em
{\it 
$^1$Harish-Chandra Research Institute \\
Chhatnag Road, Jhunsi, Allahabad 211019, India\\
$^2$Graduate School of Mathematics \\
Nagoya University, Nagoya, 464-8602, Japan \\
}
\end{center}
\vskip3cm

\begin{abstract}
We study refined B-model via the beta ensemble of matrix models. Especially, for four dimensional ${\cal N}=2$ $SU(2)$ supersymmetric gauge theories with $N_f=0,1$ and $2$ fundamental flavors, we discuss the correspondence between deformed disk amplitudes on each Seiberg-Witten curve and the Nekrasov-Shatashvili limit of the corresponding irregular one point block of a degenerate operator via the AGT correspondence. We also discuss the relation between deformed annulus amplitudes and the irregular two point block of the degenerate operator, and check a desired agreement for $N_f=0$ and $1$ cases.
\end{abstract}
\end{titlepage}


\renewcommand{\thefootnote}{\arabic{footnote}} \setcounter{footnote}{0}

\section{Introduction}

The AGT correspondence between four dimensional ${\cal N}=2$ supersymmetric gauge theory and two dimensional conformal field theory gives us several new insights into both theories. This correspondence was originally found, or conjectured by Alday, Gaiotto and Tachikawa in 2009 \cite{Alday:2009aq} as $SU(2)$ superconformal quiver gauge theory in the Omega background / Liouville field theory (Virasoro algebra) correspondence by compactifying the six dimensional $A_1$ $(2,0)$ theory on a Riemann surface with punctures \cite{Gaiotto:2009we}. The Omega background has two deformation parameters $\epsilon_1$ and $\epsilon_2$ which generate the rotation of ${\IR}^4\simeq{\IC}^2\ni (z_1,z_2)$ \cite{Nekrasov:2002qd} (see also \cite{Nekrasov:2010ka}):
\begin{equation}
(z_1, z_2) \mapsto (e^{i\epsilon_1}z_1, e^{i\epsilon_2}z_2).
\label{omega}
\end{equation}
After that, various extensions and generalizations have been discussed. For instance,
\begin{itemize}
 \item The extension to $SU(N)$ superconformal quiver gauge theory / $A_{N-1}$ Toda field theory ($W_{N-1}$-algebra) correspondence \cite{Wyllard:2009hg}, and the generalization to the non-conformal cases by decoupling the flavors \cite{Gaiotto:2009ma, Marshakov:2009gn, Taki:2009zd}. We correctively call these correspondences ``AGT correspondence''.
 \item The extension to a half BPS ``simple type'' surface operator which brakes the gauge symmetry as $SU(N)$ to $SU(N-1)\times U(1)$ / a degenerate operator in Toda field theory correspondence \cite{Alday:2009fs, Kozcaz:2010af, Dimofte:2010tz}. This correspondence is called ``AGGTV correspondence''.
\end{itemize}
In \cite{Dijkgraaf:2009pc}, Dijkgraaf and Vafa explained the AGT correspondence by topological B-model / matrix model correspondence, and by free field representation of the matrix model. In this paper, we only concentrate on $SU(2)$ cases. Although the corresponding Penner type matrix model which has logarithmic potential plays an important role in their explanation, in section 2 we consider the beta ensemble of matrix models with a polynomial potential $V(z)$ for avoiding some subtleties:
\begin{equation}
Z_{\beta}:=\frac{1}{N!(2\pi)^N}\int_{\mathbb{R}} \prod_{i=1}^Ndz_i|\Delta(z)|^{2\beta}e^{-\frac{2\sqrt{\beta}}{g_s}\sum_{i=1}^NV(z_i)},\quad \Delta(z):=\prod_{i<j}(z_i-z_j).
\label{matrix}
\end{equation}
When $\beta=1$, this model reduces to the hermitian one-matrix model. From this model, one can obtain ``refined'' topological recursion (\ref{loop_eq}) as the loop equation \cite{Chekhov:2006rq, Chekhov:2010xj, Brini:2010fc} which reduces to the Eynard-Orantin recursion \cite{Eynard:2007kz} in $\beta=1$ case.

In \cite{Dijkgraaf:2009pc}, they also proposed a refinement of topological strings by the beta deformation of matrix models. After forgetting the matrix model origin of the ``refined'' topological recursion, we utilize this recursion as the refinement of topological B-model on the Seiberg-Witten curve of Gaiotto form \cite{Alday:2009aq, Gaiotto:2009ma}. And then we study the relation between refined open topological string free energy (amplitude) and chiral block with a degenerate operator $\Phi_{1,2}$ corresponding to the ``simple type'' surface operator. This work is a refined version of the B side computation discussed in \cite{Kozcaz:2010af, Awata:2010bz} for $SU(2)$ cases.\footnote{A $U(1)$ case which has the dual description by the Penner type matrix model was discussed in \cite{Brini:2010fc}.}

In section 3, we study an operator in the beta ensemble corresponding to the degenerate operator on the CFT side as \cite{Marshakov:2010fx}, and formally discuss a relation between its correlator and ``deformed free energy'' with a deformation index $\ell$ obtained from the ``refined'' topological recursion. Especially, by taking the Nekrasov-Shatashvili (NS) limit $\epsilon_2\to 0$ \cite{Nekrasov:2009rc}, we obtain the simple relation (\ref{NS_rel}) which claims that the NS limit of the $m$ point block of the degenerate operator equals to the summation of deformed disk amplitudes on Seiberg-Witten curve obtained from the ``refined'' topological recursion.\footnote{In \cite{Nekrasov:2009rc}, it was proposed that the NS limit gives the quantization of the classical integrable system / Seiberg-Witten theory \cite{Gorsky:1995zq, Donagi:1995cf} with the ``Planck constant'' $\hbar=\epsilon_1$. In \cite{Mironov:2009uv}, it was also proposed that under this limit, the Nekrasov partition function is constructed from the Bohr-Sommerfeld period. These proposals were further studied in \cite{Mironov:2009dv, Mironov:2009ib, Popolitov:2010bz, Nekrasov:2010ka, Teschner:2010je, He:2010xa, Maruyoshi:2010iu, Marshakov:2010fx, Poghossian:2010pn, Tai:2010ps, Piatek:2011tp, Nekrasov:2011bc, Fucito:2011pn, Zenkevich:2011zx, Dorey:2011pa, Chen:2011sj, Bonelli:2011na}. The NS limit has been also discussed from the viewpoint of refined topological strings \cite{Aganagic:2011mi, Huang:2011qx}.} By direct computation of the deformed disk amplitudes, in section 4, we check the relation for $SU(2)$ theories with $N_f=0,1$ and $2$ fundamental flavors. In section 5, we discuss the deformed annulus amplitudes, and obtain the expected relation (\ref{NS_rel21s}) to the two point block of the degenerate operator, by taking the ambiguity independent parts on both sides. For $SU(2)$ theories with $N_f=0$ and $1$ fundamental flavor, we check this relation only up to $\ell=1$. Section 6 is devoted to the conclusion of this paper, and a future direction. In appendix A, we summarize the computation on the CFT side \cite{Awata:2010bz}. In appendix B, we discuss that the Bergman kernel which defines the annulus amplitude on genus one Seiberg-Witten curve can be written as a functional of the period of the curve. Appendix C is the summary of some detailed computation on the B-model side in section 4 and 5.

\section{Topological recursion of the beta ensemble}

In this section, we summarize the beta ensemble of matrix models defined by (\ref{matrix}) and its topological recursion. The connected correlator of the ``trace'' type operator,
\begin{equation}
W_h(p_1,\ldots,p_h):=\beta^{\frac{h}{2}}\bigg<\prod_{j=1}^h\sum_{i=1}^N\frac{dp_j}{p_j-z_i}\bigg>^{\scriptsize(\mbox{c})},
\end{equation}
can be expanded under $g_s \to 0, N \to \infty$, and fixed $g_sN$ as
\begin{equation}
W_h(p_1,\ldots,p_h)=\sum_{g,\ell=0}^{\infty}g_s^{2g-2+h+\ell}\gamma^{\ell}W^{(g,h)}_{\ell}(p_1,\ldots,p_h),\quad \gamma=\sqrt{\beta}-\frac{1}{\sqrt{\beta}}.
\label{Wexpand}
\end{equation}
On topological string theory side, $g$ and $h$ are interpreted as ``genus'' and ``boundary'' of the world sheet respectively, and in this beta ensemble one has new index $\ell$. In the $\beta=1$ case, the perturbative correlators $W^{(g,h)}_{\ell\geq 1}(p_1,\ldots,p_h)$ vanish, and we call these quantities ``deformed correlators'' with the deformation index $\ell$. It is well known that the (undeformed) disk and annulus correlators are given by \cite{Migdal:1984gj, Ambjorn:1992gw, Akemann:1996zr, Eynard:2004mh},
\begin{eqnarray}
\label{disc}
&&
W^{(0,1)}_{0}(p)=\oint_{\cal A}\frac{dq}{2\pi i}\frac{V'(q)}{p-q}\sqrt{\frac{\sigma(p)}{\sigma(q)}}dp=:V'(p)dp-y(p)dp,\\
&&
W^{(0,2)}_{0}(p_1,p_2)=B(p_1,p_2)-\frac{dp_1dp_2}{(p_1-p_2)^2}.
\end{eqnarray}
The counterclockwise cycle ${\cal A}=\bigcup_{i=1}^s{\cal A}_i$ is surrounding the compact support ${\cal C}$ of the eigenvalue density $\rho(z)=\lim_{N \to \infty}\frac{1}{N}\sum_{i=1}^{N}\delta(z-z_i)$ in (\ref{matrix}). This support is divided into $s$ branch cuts ${\cal C}=\bigcup_{i=1}^s{\cal C}_i$ of the spectral curve $y(p)=M(p)\sqrt{\sigma(p)}$ defined by (\ref{disc}). When the potential $V(z)$ is a polynomial, we see that $M(p)$ is a rational function, and $\sigma(p)=p^{2s-1}+\cdots$, or $p^{2s}+\cdots$ is a polynomial. $B(p_1,p_2)$ is the Bergman kernel on the spectral curve. The other correlators $W^{(g,h)}_{\ell}(p_H)=\widetilde{W}^{(g,h)}_{\ell}(p_H)~\mbox{for}~(g,h,\ell)\neq(0,1,0),(0,2,0)$ are obtained from the ``refined'' topological recursion \cite{Brini:2010fc, Chekhov:2006rq, Chekhov:2010xj}:
\begin{eqnarray}
&&
\widetilde{W}^{(g,h+1)}_{\ell}(p,p_H)=\oint_{\cal A}\frac{1}{2\pi i}\frac{dE_q(p)}{y(q)dq}\bigg\{\widetilde{W}^{(g-1,h+2)}_{\ell}(q,q,p_H)\nonumber\\
&&
\hspace{3.3em}+\sum_{k=0}^{g}\sum_{n=0}^{\ell}\sum_{\emptyset=J\subseteq H}\widetilde{W}^{(g-k,|J|+1)}_{\ell-n}(q,p_J)\widetilde{W}^{(k,|H|-|J|+1)}_{n}(q,p_{H \backslash J})+dq\frac{d}{dq}\widetilde{W}^{(g,h+1)}_{\ell-1}(q,p_H)\bigg\},\nonumber\\
&&
\widetilde{W}^{(0,1)}_{0}(p)=0,\quad \widetilde{W}^{(0,2)}_{0}(p_1,p_2)=B(p_1,p_2)-\frac{dp_1dp_2}{2(p_1-p_2)^2},\quad \frac{d}{dp}\widetilde{W}^{(0,1)}_{0}(p)=\frac{d}{dp}W^{(0,1)}_{0}(p),\nonumber\\
&&
\label{loop_eq}
\end{eqnarray}
where $H=\{1,2,\ldots,h\}\supset J=\{i_1,i_2,\ldots,i_j\},H\backslash J=\{i_{j+1},i_{j+2},\ldots,i_h\}$. When $\beta=1$, the last term of this recursion vanishes, and then this reduces to the Eynard-Orantin recursion \cite{Eynard:2007kz, Alexandrov:2003pj, Eynard:2004mh}. The third type differential $dE_q(p)$ is given as \cite{Eynard:2004mh},
\begin{equation}
dE_q(p)=\frac{\sqrt{\sigma(q)}}{2\sqrt{\sigma(p)}}\Big(\frac{1}{p-q}-\sum_{i=1}^{s-1}C_i(q)L_i(p)\Big)dp,\quad C_i(q)=\frac{1}{2\pi i}\oint_{q\not\in {\cal A}_i}\frac{d\widetilde{q}}{(\widetilde{q}-q)\sqrt{\sigma(\widetilde{q})}}.
\end{equation}
If $q$ is contained in the cycle ${\cal A}_i$, then $C_i(q)$ must be replaced with $C_i^{\scriptsize\mbox{reg}}(q)=C_i(q)\pm\frac{1}{\sqrt{\sigma(q)}}$, where the sign depends on the branch. $L_i(p)$ are the (normalized) bases of the holomorphic differentials on the spectral curve:
\begin{equation}
\frac{1}{2\pi i}\oint_{{\cal A}_j}\frac{L_i(p)}{\sqrt{\sigma(p)}}dp=\delta_{i,j},\quad L_i(p)=\sum_{j=1}^{s-1}L_{j,i}p^{j-1},\quad i=1,\ldots,s-1.
\end{equation}

In the two-cut case, the differential
\begin{equation}
\frac{dE_q(p)}{y(q)}=\frac{1}{2\sqrt{\sigma(p)}M(q)}\Big(\frac{1}{p-q}-LC^{\scriptsize\mbox{(reg)}}(q)\Big)dp
\label{dEq}
\end{equation}
is concretely written by the complete elliptic integral of the first and third kind as \cite{Bouchard:2007ys},
\begin{eqnarray}
&&
L:=L_1=-\frac{\pi \sqrt{(q_1-q_3)(q_2-q_4)}}{2K(k)},\nonumber\\
&&
LC^{\scriptsize\mbox{reg}}(q)=\frac{(q_2-q_3)\Pi(n_1,k)}{(q-q_2)(q-q_3)K(k)}-\frac{1}{q-q_2},\quad LC(q)=\frac{(q_3-q_2)\Pi(n_4,k)}{(q-q_2)(q-q_3)K(k)}-\frac{1}{q-q_3},\nonumber\\
&&
k^2=\frac{(q_1-q_2)(q_3-q_4)}{(q_1-q_3)(q_2-q_4)},\quad n_1=\frac{(q_4-q_3)(q-q_2)}{(q_4-q_2)(q-q_3)},\quad n_4=\frac{(q_2-q_1)(q-q_3)}{(q_3-q_1)(q-q_2)},\nonumber\\
&&
K(k)=\int_0^1\frac{dt}{\sqrt{(1-t^2)(1-k^2t^2)}},\quad \Pi(n,k)=\int_0^1\frac{dt}{(1-nt^2)\sqrt{(1-t^2)(1-k^2t^2)}},
\label{dEqL}
\end{eqnarray}
where $q_1 < q_2 < q_3 < q_4$ are the branch points on the spectral curve. The differential (\ref{dEq}) is defined by $C^{\scriptsize\mbox{reg}}(q)$ ($C(q)$) for $q$ inside (outside) the cycle ${\cal A}_1$ around the cut ${\cal C}_1=[q_1,q_2]$.

\section{The Nekrasov-Shatashvili limit of refined B-model}

Here, we discuss the NS limit of the beta ensemble (\ref{matrix}). We consider the theory which is defined by the topological recursion (\ref{loop_eq}) on spectral curve as refined topological B-model on spectral curve.

To discuss the NS limit of the beta ensemble, let us introduce the parameters
\begin{equation}
\epsilon_1=-\sqrt{\beta}g_s,\quad \epsilon_2=\frac{g_s}{\sqrt{\beta}}.
\label{matparam}
\end{equation}
These parameters are identified with the Omega background parameters in (\ref{omega}). And then two types of NS limit are defined \cite{Nekrasov:2009rc}:
\begin{eqnarray}
\label{limit A}
&&
\mbox{limit A}:\quad \epsilon_2\to0,~\epsilon_1:~\mbox{finite},\quad(g_s\to 0,~\beta\to \infty),\\
&&
\mbox{limit B}:\quad \epsilon_1\to0,~\epsilon_2:~\mbox{finite},\quad(g_s\to 0,~\beta\to 0).
\label{limit B}
\end{eqnarray}

We now consider the two ``determinant'' type operators \cite{Aganagic:2011mi},
\begin{equation}
{\cal O}_{\alpha}(p):=e^{\frac{1}{\epsilon_{\alpha}}V(p)}\prod_{i=1}^N(p-z_i)^{\frac{\epsilon_1}{\epsilon_{\alpha}}},\quad \alpha=1,2.
\label{det_op}
\end{equation}
By using the free chiral boson representation for the beta ensemble \cite{Kostov:1999xi, Aganagic:2003qj, Marshakov:2010fx, Aganagic:2011mi}:
\begin{equation}
\phi(p)=\frac{1}{g_s}V(p)-\frac{g_s}{\epsilon_2}\sum_{i=1}^N\log(p-z_i),
\label{chil_bos}
\end{equation}
the operator ${\cal O}_{\alpha}(p)$ is represented as free fermion ${\cal O}_{\alpha}(p)=\exp\big(\frac{g_s}{\epsilon_{\alpha}}\phi(p)\big)$, and from the correspondence to the B-brane in the topological B-model this operator is called ``$\epsilon_{\alpha}$-brane'' \cite{Aganagic:2011mi}. Let us consider the $m$ point correlator of the ``$\epsilon_{\alpha}$-branes'':
\begin{eqnarray}
\left<{\cal O}_{\alpha}(p_1)\cdots{\cal O}_{\alpha}(p_m)\right>^{\scriptsize(\mbox{c})}-\log Z_{\beta}&=&
\sum_{h=1}^{\infty}\frac{1}{h!}\Big(\frac{g_s}{\epsilon_{\alpha}}\Big)^h\big<\big(\phi(p_1)+\cdots+\phi(p_m)\big)^h\big>^{\scriptsize(\mbox{c})}\nonumber\\
\label{mult_det}
&=&
\sum_{h=1}^{\infty}\frac{1}{h!}\Big(-\frac{g_s}{\epsilon_{\alpha}}\Big)^h\sum_{i_1,\ldots,i_h=1}^m
{\cal F}^{(h)}(p_{i_1},\ldots,p_{i_h}),
\end{eqnarray}
where ${\cal F}^{(h)}(p_1,\ldots,p_h):=
(-1)^h\big<\phi(p_1)\cdots\phi(p_h)\big>^{\scriptsize(\mbox{c})}$. By (\ref{chil_bos}), one can find that the right hand side of (\ref{mult_det}) is rewritten as
\begin{equation}
\frac{1}{\epsilon_{\alpha}}\big(V(p_1)+\cdots+V(p_m)\big)+
\sum_{h=1}^{\infty}\frac{1}{h!}\Big(-\frac{g_s}{\epsilon_{\alpha}}\Big)^h\sum_{i_1,\ldots,i_h=1}^m\int^{p_{i_1}}\cdots\int^{p_{i_h}}W_h(p_{i_1}',\ldots,p_{i_h}'),
\end{equation}
and then the following perturbative expansion of (\ref{mult_det}) is obtained \cite{Marshakov:2010fx}:
\begin{eqnarray}
&&
\sum_{i,j=0}^{\infty}\frac{\epsilon_1^i\epsilon_2^j}{\epsilon_{\alpha}}{\cal S}^{(m)}_{\alpha;i,j}(p_1,\ldots,p_m)\hspace{25em}\nonumber\\
\label{F_exp}
&&\hspace{4em}
=\sum_{h=1,g,\ell=0}^{\infty}(-1)^{g-1+\ell}\frac{(\epsilon_1\epsilon_2)^{g-1+h}}{\epsilon_{\alpha}^h}(\epsilon_1+\epsilon_2)^{\ell}\frac{1}{h!}\sum_{i_1,\ldots,i_h=1}^m{\cal F}^{(g,h)}_{\ell}(p_{i_1},\ldots,p_{i_h}),\\
&&
{\cal F}^{(g,h)}_{\ell}(p_1,\ldots,p_h):=\int^{p_1}\cdots\int^{p_h}
W^{(g,h)}_{\ell}(p_1',\ldots,p_h')-V(p_1)\delta_{g,0}\delta_{h,1}\delta_{\ell,0},
\label{Fenergy}
\end{eqnarray}
where ${\cal S}^{(m)}_{\alpha;i,j}(p)$ are the expansion coefficients of the left hand side of (\ref{mult_det}). On the right hand side of (\ref{F_exp}), we have used the expansion (\ref{Wexpand}), and defined the perturbative free energies ${\cal F}^{(g,h)}_{\ell}(p_1,\ldots,p_h)$ by (\ref{Fenergy}). By this definition, by absorbing the potential term $V(p)$ of (\ref{disc}) we redefine the disk correlator as \cite{Marshakov:2010fx},
\begin{equation}
(\ref{disc})\quad\longrightarrow\quad
W^{(0,1)}_{0}(p)=-y(p)dp.
\label{redef01}
\end{equation}
Actually if $V''(p)$ is an analytic function inside ${\cal A}$, the higher order correlators are not changed under this redefinition as seen from the recursion (\ref{loop_eq}).

Let us consider the ``$\epsilon_1$, (or $\epsilon_2$)-brane'', and by taking the limit A (\ref{limit A}), (or the limit B (\ref{limit B})), we obtain the relation:
\begin{equation}
{\cal S}^{(m)}_{1;\ell,0}(p_1,\ldots,p_m)={\cal S}^{(m)}_{2;0,\ell}(p_1,\ldots,p_m)=(-1)^{\ell+1}\sum_{i=1}^m{\cal F}^{(0,1)}_{\ell}(p_i)
\label{NS_rel}
\end{equation}
as the coefficients of $\epsilon_1^{\ell-1}$, (or $\epsilon_2^{\ell-1}$). From this relation, we see that under the NS limit, the $m$ point correlator of the ``$\epsilon_{\alpha}$-brane'' is factorized into the one point correlators:
\begin{equation}
{\cal S}^{(m)}_{1;\ell,0}(p_1,\ldots,p_m)={\cal S}^{(m)}_{2;0,\ell}(p_1,\ldots,p_m)=\sum_{i=1}^m{\cal S}^{(1)}_{1;\ell,0}(p_i)=\sum_{i=1}^m{\cal S}^{(1)}_{2;0,\ell}(p_i).
\label{NS_rel2}
\end{equation}
In the next section, we compute the right hand side of (\ref{NS_rel}). By the recursion (\ref{loop_eq}) and the redefinition (\ref{redef01}), the $\ell$-th order deformed disk correlator $W^{(0,1)}_{\ell\geq 1}(p)=\widetilde{W}^{(0,1)}_{\ell\geq 1}(p)$ is obtained from the planar ``refined'' topological recursion:
\begin{eqnarray}
&&
\widetilde{W}^{(0,1)}_{\ell}(p)=
\oint_{\cal A}\frac{1}{2\pi i}\frac{dE_q(p)}{y(q)dq}\bigg\{\sum_{n=0}^{\ell}\widetilde{W}^{(0,1)}_{\ell-n}(q)\widetilde{W}^{(0,1)}_{n}(q)+dq\frac{d}{dq}\widetilde{W}^{(0,1)}_{\ell-1}(q)\bigg\},\nonumber\\
&& \widetilde{W}^{(0,1)}_{0}(p)=0,\quad \frac{d}{dp}\widetilde{W}^{(0,1)}_{0}(p)=\frac{d}{dp}W^{(0,1)}_{0}(p),\quad W^{(0,1)}_{0}(p)=-y(p)dp.
\label{1loop_eq}
\end{eqnarray}

\section{Deformed disk amplitudes}

In this section, we check the relation (\ref{NS_rel}) for $m=1$ case. For the direct check we consider the $SU(2)$ gauge theory with the $N_f=0,1$ and $2$ fundamental flavors. And then, via free field realization of the beta ensemble (\ref{matrix}), the two ``determinant'' type operators (\ref{det_op}) correspond to the degenerate primary operators $\Phi_{1,2}$ with the Liouville momentum $-\frac{1}{2b}$, and $\Phi_{2,1}$ with the Liouville momentum $-\frac{b}{2}$ in the Liouville CFT \cite{Marshakov:2010fx, Aganagic:2011mi}:
\begin{equation}
{\cal O}_{1}(p)\longleftrightarrow \Phi_{1,2}(p),\quad {\cal O}_{2}(p)\longleftrightarrow \Phi_{2,1}(p).
\label{det_deg}
\end{equation}
Using this correspondence, we compute the left hand side in the relation (\ref{NS_rel}) as the perturbative expansion of the correlation function of the degenerate operator $\Phi_{1,2}$. By taking the appropriate CFT vacuum $|G(a)\rangle$ \cite{Gaiotto:2009ma} corresponding to each $SU(2)$ gauge theory with the Coulomb moduli parameter $a$ via the AGT correspondence, we consider the perturbative expansion of the chiral one point block:
\begin{equation}
\log\frac{\langle G(a-\frac{1}{4b})|\Phi_{1,2}(p)|G(a+\frac{1}{4b})\rangle}{\langle G(a)|G(a)\rangle}=:\sum_{i,j=0}^{\infty}\frac{\varepsilon_1^i\varepsilon_2^j}{\varepsilon_1}{\cal G}^{(1)}_{i,j}(p),
\label{Gaiopen}
\end{equation}
where after introducing the mass scale $g_s$ on the CFT side as the scaling $a \to a/g_s$, we have introduced the parameters
\begin{equation}
\varepsilon_1=b g_s,\quad \varepsilon_2=\frac{g_s}{b}.
\label{varep}
\end{equation}
For the charge conservation, the momentum $a$ of the left hand side in (\ref{Gaiopen}) is shifted. These parameters correspond to (\ref{matparam}) introduced on the matrix model side, and then the free energies ${\cal G}^{(1)}_{i,j}(p)$ correspond to ${\cal S}^{(1)}_{1;i,j}(p)$. In this correspondence, we have identified parameters as $\varepsilon_1=\epsilon_1$ and $\varepsilon_2=\epsilon_2$. Therefore, one can expect the relation:
\begin{equation}
{\cal G}^{(1)}_{\ell,0}(p)=(-1)^{\ell+1}{\cal F}^{(0,1)}_{\ell}(p).
\label{NS_relCFT}
\end{equation}
Note that the left hand side of this relation gives us the momentum shift invariant solutions in (\ref{Gaiopen}). The computations on these CFT side are summarized in appendix A.

The ``disk'' part of the relation (\ref{NS_relCFT}):
\begin{equation}
{\cal G}^{(1)}_{0,0}(p)=-{\cal F}^{(0,1)}_{0}(p)=\int^py(p')dp'
\label{disk_rel}
\end{equation}
is generally proved on the CFT side \cite{Alday:2009fs}. Note that for the cases of $N_f=0$ and $1$, the matrix model, or beta ensemble realizations as (\ref{matrix}) are not known, but we can consider the refined B-model on each corresponding Seiberg-Witten curve by the recursion (\ref{loop_eq}).\footnote{In \cite{Eguchi:2009gf}, the Penner type matrix models corresponding to the $SU(2)$ gauge theories with $N_f=2$ and $3$ fundamental flavors were given by decoupling the flavors of the Penner type matrix model for $N_f=4$ fundamental flavors proposed in \cite{Dijkgraaf:2009pc}.}

In case that the spectral curve $y(p)=M(p)\sqrt{\sigma(p)}$ has two cuts ${\cal C}_1=[q_1,q_2]$ and ${\cal C}_2=[q_3,q_4]$, using (\ref{dEq}), from the planar recursion (\ref{1loop_eq}), the first order deformed disk correlator $W^{(0,1)}_{1}(p)$ is given by \cite{Brini:2010fc},
\begin{eqnarray}
&&
W^{(0,1)}_{1}(p)=W^{(0,1)}_{1,A}(p)+W^{(0,1)}_{1,B}(p),\nonumber\\
&&
W^{(0,1)}_{1,A}(p):=\frac{dp}{4\pi i\sqrt{\sigma(p)}}\oint_{\cal A}\frac{y'(q)dq}{(q-p)M(q)},\nonumber\\
&&
W^{(0,1)}_{1,B}(p):=\frac{dp}{4\pi i\sqrt{\sigma(p)}}\Big[\oint_{{\cal A}_1}LC^{\scriptsize\mbox{reg}}(q)+\oint_{{\cal A}_2}LC(q)\Big]\frac{y'(q)dq}{M(q)},
\label{disk0_cor}
\end{eqnarray}
where ${\cal A}_1$ and ${\cal A}_2$ are the cycles surrounding the cuts ${\cal C}_1$ and ${\cal C}_2$ respectively. In this section, we check the relation (\ref{NS_relCFT}) up to the second order deformed disk correlator
\begin{eqnarray}
&&
W^{(0,1)}_{2}(p)=W^{(0,1)}_{2,A}(p)+W^{(0,1)}_{2,B}(p),\nonumber\\
&&
W^{(0,1)}_{2,A}(p):=-\frac{dp}{4\pi i\sqrt{\sigma(p)}}\oint_{\cal A}\frac{1}{(q-p)M(q)dq}\Big[W^{(0,1)}_{1}(q)^2+dq\frac{d}{dq}W^{(0,1)}_{1}(q)\Big],\nonumber\\
&&
W^{(0,1)}_{2,B}(p):=-\frac{dp}{4\pi i\sqrt{\sigma(p)}}\Big[\oint_{{\cal A}_1}LC^{\scriptsize\mbox{reg}}(q)+\oint_{{\cal A}_2}LC(q)\Big]\frac{1}{M(q)dq}\Big[W^{(0,1)}_{1}(q)^2+dq\frac{d}{dq}W^{(0,1)}_{1}(q)\Big].\nonumber\\
&&
\label{disk1_cor}
\end{eqnarray}
The detailed computations are summarized in appendix C.

\subsection{$N_f=0$ case}

For the pure $SU(2)$ case with the quantum Coulomb moduli parameter $u$, and the dynamical scale parameter $\Lambda$, the Seiberg-Witten curve is given by \cite{Gaiotto:2009ma},
\begin{equation}
y(p)=M(p)\sqrt{\sigma(p)},\quad \sigma(p)=-p\Big(p^2-\frac{u}{\Lambda^2}p+1\Big),~M(p)=\frac{\Lambda}{p^2}.
\label{SW_curve0}
\end{equation}
The branch points of this curve are
\begin{equation}
q_1=0,\quad q_2=\frac{u-\sqrt{u^2-4\Lambda^4}}{2\Lambda^2},\quad q_3=\frac{u+\sqrt{u^2-4\Lambda^4}}{2\Lambda^2}=q_2^{-1},\quad q_4=\infty.
\end{equation}
By computing the period of this curve,
\begin{equation}
\frac{da(u)}{du}=\oint_{{\cal A}_1} \frac{dp}{2\pi i}\frac{\partial y(p)}{\partial u}
=\frac{i}{2\pi \Lambda}\int_{q_1}^{q_2} \frac{dp}{\sqrt{\sigma(p)}}=\frac{1}{\pi \Lambda \sqrt{q_3-q_1}}K(k),\quad k^2=\frac{q_2-q_1}{q_3-q_1},
\end{equation}
one obtains the weak coupling expansion of the quantum modulus $u$,
\begin{equation}
u(a)=a^2+\frac{1}{2a^2}\Lambda^4+\frac{5}{32a^6}\Lambda^8+\frac{9}{64a^{10}}\Lambda^{12}
+\frac{1469}{8192a^{14}}\Lambda^{16}+\frac{4471}{16384a^{18}}\Lambda^{20}+{\cal O}(\Lambda^{24}).
\end{equation}

By computing the disk amplitude ${\cal F}^{(0,1)}_0(p)$, we have directly checked that this coincides with $-{\cal G}^{(1)}_{0,0}(p)$ computed in (\ref{F00}) up to $\Lambda^{12}$ \cite{Awata:2010bz}, and then the relation (\ref{disk_rel}) surely holds. Next, let us compute the first order deformed disk amplitude ${\cal F}^{(0,1)}_1(p)$ by (\ref{disk0_cor}). As discussed in appendix C.1, we can show $W^{(0,1)}_{1,B}(p)=0$, and then
\begin{equation}
{\cal F}^{(0,1)}_1(p)=\int^pW^{(0,1)}_{1}(p')=-\frac12\int^p\frac{y'(p')}{y(p')}dp'=-\frac12\log y(p)+c
\end{equation}
is obtained. By taking $c=\frac12 \log a$, we have checked that this result coincides with ${\cal G}^{(1)}_{1,0}(p)$ computed in (\ref{F01}) up to $\Lambda^{12}$. The second order deformed disk amplitude ${\cal F}^{(0,1)}_2(p)$ is also computed by (\ref{disk1_cor}),
\begin{eqnarray}
&&
{\cal F}^{(0,1)}_2(p)=\int^p\bigg\{
\frac{V_1^{(0,1)}(p')}{2y(p')dp'}
-\frac{dp'}{2\sqrt{\sigma(p')}}\Big[\oint_{{\cal A}_1}LC^{\scriptsize\mbox{reg}}(q)+\oint_{{\cal A}_2}LC(q)\Big]\frac{1}{2\pi i}\frac{V_1^{(0,1)}(q)}{M(q)dq}\bigg\},\nonumber\\
&&
V_1^{(0,1)}(p):=W^{(0,1)}_{1}(p)^2+dp\frac{d}{dp}W^{(0,1)}_{1}(p),\quad W^{(0,1)}_{1}(p)=-\frac{y'(p)}{2y(p)}dp,
\end{eqnarray}
and we have checked that this coincides with $-{\cal G}^{(1)}_{2,0}(p)$ computed in (\ref{F02}) up to $\Lambda^{12}$.

\subsection{$N_f=1$ case}

The Seiberg-Witten curve for the $SU(2)$ gauge theory with one fundamental matter of mass $m$ is \cite{Gaiotto:2009ma},
\begin{equation}
y(p)=M(p)\sqrt{\sigma(p)},\quad \sigma(p)=p\big(p^3+\frac{2m}
{\Lambda}p^2+\frac{u}{\Lambda^2}p-1\big),~M(p)=\frac{\Lambda}{p^2}.
\label{SW_curve1}
\end{equation}
The branch points of this curve are perturbatively obtained as
\begin{equation}
q_{1,2}=-\frac{m \pm \sqrt{m^2-u}}{\Lambda}+\frac{\Lambda^2}{2(m^2\pm m\sqrt{m^2-u}-u)}+{\cal O}(\Lambda^5),\quad q_3=0,\quad q_4=\frac{\Lambda^2}{u}+{\cal O}(\Lambda^5),
\end{equation}
where for the double signs, $q_1$ takes the plus signs, and $q_2$ takes the minus signs. Because $q_1$ and $q_2$ are expanded around $\infty$, we need to take the cycle ${\cal A}_1$ as containing the point at infinity. By the period computation
\begin{equation}
\frac{da(u)}{du}=\frac{1}{\pi \Lambda \sqrt{(q_1-q_3)(q_2-q_4)}}K(k),\quad k^2
=\frac{(q_1-q_2)(q_3-q_4)}{(q_1-q_3)(q_2-q_4)},
\end{equation}
one obtains 
\begin{equation}
u(a)=a^2-\frac{m}{a^2}\Lambda^3-\frac{3a^2-5m^2}{8a^6}\Lambda^6+{\cal O}(\Lambda^9).
\end{equation}
As in the pure case, by comparing with (\ref{F10}) -- (\ref{F12}), we have checked the expected relation (\ref{NS_relCFT}) for $\ell=0,1$ and $2$ up to $\Lambda^6$. Especially $W^{(0,1)}_{1,B}(p)=0$ is proved in appendix C.1, and we checked the same relation to the $N_f=0$ case:
\begin{equation}
{\cal G}^{(1)}_{1,0}(p)={\cal F}^{(0,1)}_1(p)=-\frac12\log \frac{y(p)}{a}.
\label{firstdisk}
\end{equation}

\subsection{$N_f=2$ case}

Finally we consider the $SU(2)$ gauge theory with two (anti-)fundamental matters of masses $m_1$ and $m_2$. The Seiberg-Witten curve is given by \cite{Gaiotto:2009ma},\footnote{The genus one amplitude ${\cal F}^{(1,0)}_0$ on this curve was discussed in \cite{Fujita:2009gf}.}
\begin{equation}
y(p)=M(p)\sqrt{\sigma(p)},\quad \sigma(p)=p^4+\frac{2m_2}{\Lambda}p^3+\frac{u}
{\Lambda^2}p^2+\frac{2m_1}{\Lambda}p+1,~M(p)=\frac{\Lambda}{p^2}.
\label{SW_curve2}
\end{equation}
The branch points of this curve are
\begin{eqnarray}
&&
q_{1,2}=-\frac{m_2\pm\sqrt{m_2^2-u}}{\Lambda}+\frac{m_1(m_2\pm\sqrt{m_2^2-u})\Lambda}{2um_2(1-m_2^2)\pm(u-2m_2^2)\sqrt{m_2^2-u}}+{\cal O}(\Lambda^3),\nonumber\\
&&
q_{3,4}=-\frac{m_1\pm\sqrt{m_1^2-u}}{u}\Lambda+{\cal O}(\Lambda^3),
\end{eqnarray}
where for the double signs, $q_1$ and $q_3$ take the plus signs, and $q_2$ and $q_4$ take the minus signs. Note that as the $N_f=1$ case, because $q_1$ and $q_2$ ($q_3$ and $q_4$) are expanded around $\infty$ (0), we need to take the cycles ${\cal A}_1$ and ${\cal A}_2$ as containing $\infty$ and $0$ respectively. The computation of the ${\cal A}_1$-period gives us
\begin{equation}
u(a)=a^2+\frac{2m_1m_2}{a^2}\Lambda^2+\frac{5m_1^2m_2^2-3(m_1^2+m_2^2)a^2+a^4}{2a^6}\Lambda^4+{\cal O}(\Lambda^6).
\end{equation}
In this case, by comparing with (\ref{F20}) -- (\ref{F22}), we have also checked the expected relation (\ref{NS_relCFT}) for $\ell=0,1$ and $2$ up to $\Lambda^4$, where as same as the $N_f=0$ and $1$ cases, $W^{(0,1)}_{1,B}(p)=0$ is proved in appendix C.1, and we checked the relation (\ref{firstdisk}).

\section{Deformed annulus amplitudes}

Here we discuss the deformed annulus amplitudes on Seiberg-Witten curve for $SU(2)$ gauge theory. For the $m$ point correlator of the ``$\epsilon_1$-branes'', by comparing the coefficients of $\epsilon_1^{\ell-1}\epsilon_2$ in (\ref{F_exp}), one obtains
\begin{equation}
{\cal S}^{(m)}_{1;\ell,1}(p_1,\ldots,p_m)=
(-1)^{\ell+1}\Big(\frac12\sum_{i,j=1}^m{\cal F}^{(0,2)}_{\ell}(p_i,p_j)-(\ell+1)\sum_{i=1}^m{\cal F}^{(0,1)}_{\ell+1}(p_i)+
\sum_{i=1}^m{\cal F}^{(1,1)}_{\ell-1}(p_i)\Big).
\label{NS_rel21}
\end{equation}
We now consider the case of $m=2$, and as in section 4, let us identify ${\cal S}^{(2)}_{1;i,j}(p_1,p_2)$ with ${\cal G}^{(2)}_{i,j}(p_1,p_2)$ defined by the perturbative expansion of the chiral two point block:
\begin{equation}
\log\frac{\langle G(a-\frac{1}{2b})|\Phi_{1,2}(p_1)\Phi_{1,2}(p_2)|G(a+\frac{1}{2b})\rangle}{\langle G(a)|G(a)\rangle}=:\sum_{i,j=0}^{\infty}\frac{\varepsilon_1^i\varepsilon_2^j}{\varepsilon_1}{\cal G}^{(2)}_{i,j}(p_1,p_2).
\label{Gaiopen2}
\end{equation}
As discussed in appendix A, we can check the factorization property (\ref{NS_rel2}): ${\cal G}^{(2)}_{i,0}(p_1,p_2)={\cal G}^{(1)}_{i,0}(p_1)+{\cal G}^{(1)}_{i,0}(p_2)$. Because the free energies ${\cal G}^{(2)}_{i,1}(p_1,p_2)$ depend on the momentum shift of in- and out-states, we need to take the shift invariant free energies $\widetilde{\cal G}^{(2)}_{i,1}(p_1,p_2)$ by considering the invariant parts under the shift
\begin{equation}
{\cal G}^{(2)}_{i,1}(p_1,p_2)+c \frac{\partial}{\partial a}{\cal G}^{(2)}_{i,0}(p_1,p_2)={\cal G}^{(2)}_{i,1}(p_1,p_2)+c\frac{\partial}{\partial a}\big({\cal G}^{(1)}_{i,0}(p_1)+{\cal G}^{(1)}_{i,0}(p_2)\big).
\label{shift_inv}
\end{equation}
For the cases in appendix A, we see that the free energies $\widetilde{\cal G}^{(2)}_{i,1}(p_1,p_2)$ have the expansion:
\begin{equation}
\widetilde{\cal G}^{(2)}_{i,1}(p_1,p_2)=\sum_{n_1,n_2\neq 0}a_{n_1,n_2}^{(i,1)}p_1^{n_1}p_2^{n_2}+a_{0,0}^{(i,1)},
\label{shift_invz}
\end{equation}
by neglecting the singular term at $p_1=p_2$. And then by (\ref{NS_rel21}), we obtain the relation:
\begin{equation}
\widetilde{\cal G}^{(2)}_{\ell,1}(p_1,p_2)-a_{0,0}^{(\ell,1)}=(-1)^{\ell+1}{\cal F}^{(0,2)}_{\ell}(p_1,p_2),
\label{NS_rel21s}
\end{equation}
where on the right hand side, we have only considered the ``universal terms'' which do not depend on the constants of integration in (\ref{Fenergy}).

In the genus one with two sheeted cases, the Bergman kernel is given by the Akemann formula \cite{Akemann:1996zr} by the complete elliptic integral of the first and second kind with the modulus $k^2$ defined in (\ref{dEqL}):
\begin{eqnarray}
\label{Bergman}
&&
B(p_1,p_2)=\frac{dp_1dp_2}{2(p_1-p_2)^2}\bigg[\frac{f(p_1,p_2)+G(k)(p_1-p_2)^2}{\sqrt{\sigma(p_1)\sigma(p_2)}}+1\bigg],\\
&&
f(p_1,p_2)=p_1^2p_2^2-\frac{S_1}{2}p_1p_2(p_1+p_2)+\frac{S_2}{6}(p_1^2+4p_1p_2+p_2^2)-\frac{S_3}{2}(p_1+p_2)+S_4,\nonumber\\
&&
G(k)=-\frac16S_2+\frac12(q_1q_2+q_3q_4)-\frac{E(k)}{2K(k)}(q_1-q_3)(q_2-q_4),\quad
E(k)=\int_0^1dt\sqrt{\frac{1-k^2t^2}{1-t^2}},\nonumber
\end{eqnarray}
where $S_d$ are the degree $d$ elementary symmetric polynomials of the branch points $q_i$ obtained from $f(p,p)=\sigma(p)=\prod_{i=1}^4(p-q_i)$.

From the recursion (\ref{loop_eq}), the first order deformed annulus amplitude ${\cal F}_1^{(0,2)}(p_1,p_2)=\int^{p_1}\int^{p_2}W_1^{(0,2)}(p_1',p_2')$ is given by
\begin{eqnarray}
&&
W_1^{(0,2)}(p_1,p_2)=W_{1,A}^{(0,2)}(p_1,p_2)+W_{1,B}^{(0,2)}(p_1,p_2),\nonumber\\
&&
W_{1,A}^{(0,2)}(p_1,p_2):=-\frac{dp_1}{4\pi i\sqrt{\sigma(p_1)}}\oint_{\cal A}\frac{V_0^{(0,2)}(q,p_2)}{(q-p_1)M(q)dq},\nonumber\\
&&
W_{1,B}^{(0,2)}(p_1,p_2):=-\frac{dp_1}{4\pi i\sqrt{\sigma(p_1)}}\Big[\oint_{{\cal A}_1}LC^{\scriptsize\mbox{reg}}(q)+\oint_{{\cal A}_2}LC(q)\Big]\frac{V_0^{(0,2)}(q,p_2)}{M(q)dq},\nonumber\\
&&
V_0^{(0,2)}(p_1,p_2):=\Big[2W_1^{(0,1)}(p_1)+dp_1\frac{d}{dp_1}\Big]\Big[B(p_1,p_2)-\frac{dp_1dp_2}{2(p_1-p_2)^2}\Big].
\label{Bergmand1}
\end{eqnarray}
In the following, for $N_f=0$ and $1$ cases, we check the relation (\ref{NS_rel21s}) up to $\ell=1$ by taking the ``universal terms'' on the right hand side.

For $N_f=0$ case (\ref{SW_curve0}), the annulus amplitude is obtained as computed in \cite{Awata:2010bz}. In this computation, because the Seiberg-Witten curve (\ref{SW_curve0}) has cubic form $\sigma(p)=-p^3+S_2p^2-S_3p+S_4$, we need to replace $f(p_1,p_2)$ and $G(k)$ in the formula (\ref{Bergman}) with
\begin{eqnarray}
&&
{\widetilde f}(p_1,p_2)=-\frac12p_1p_2(p_1+p_2)+\frac{S_2}{6}(p_1^2+4p_1p_2+p_2^2)-\frac{S_3}{2}(p_1+p_2)+S_4,\\
&&
{\widetilde G}(k)=-\frac16S_2+\frac12 q_3-\frac{E(k)}{2K(k)}(q_3-q_1),\quad k^2=\frac{q_2-q_1}{q_3-q_1}.
\end{eqnarray}
By comparing this result with (\ref{FF00}), we have checked the relation (\ref{NS_rel21s}) for $\ell=0$ up to $\Lambda^8$ \cite{Awata:2010bz}. By (\ref{Bergmand1}), the first order deformed annulus amplitude ${\cal F}_1^{(0,2)}(p_1,p_2)$ can be also computed, and by comparing with (\ref{FF01}), 
we have checked the relation (\ref{NS_rel21s}) for $\ell=1$ up to $\Lambda^8$.

For $N_f=1$ case (\ref{SW_curve1}), as above, the (deformed) annulus amplitudes ${\cal F}_0^{(0,2)}(p_1,p_2)$ and ${\cal F}_1^{(0,2)}(p_1,p_2)$ are also computed, and by comparing these results with (\ref{FF10}) and (\ref{FF11}) respectively, we have checked the relation (\ref{NS_rel21s}) up to $\Lambda^4$.\footnote{For $\ell=0$, it was checked in \cite{Awata:2010bz}.}

As discussed in this section, for the higher deformed planar topological open string amplitudes, by comparing the coefficients of $\epsilon_1^{\ell-1}\epsilon_2^{h-1}$ in (\ref{F_exp}), and taking the ``universal terms'' as in (\ref{NS_rel21s}), we can also expect more general relation:
\begin{equation}
\widetilde{\cal G}^{(m)}_{\ell,h-1}(p_1,\ldots,p_m)-a_{0,0}^{(\ell,h-1)}=(-1)^{\ell+1}\sum_{1\le i_1 < \cdots < i_h \le m}{\cal F}^{(0,h)}_{\ell}(p_{i_1},\ldots,p_{i_h}),\quad 1\leq h \leq m,
\label{NS_relgen}
\end{equation}
to the expansion of the chiral $m$ point block defined as (\ref{Gaiopen2}) and (\ref{shift_invz}).

\section{Conclusion and future direction}

In this paper, we have studied a refinement of topological B-model via beta ensemble of matrix models. Especially, we confirmed the agreement between the deformed planar open B-model amplitudes on Seiberg-Witten curve and chiral blocks with the degenerate operator $\Phi_{1,2}$ for some $SU(2)$ cases which was inspired from the AGT correspondence. Although there are some ambiguities on both theories: ambiguities of the constants of integration (\ref{Fenergy}) on the B-model side, and ambiguities of the momentum sift of in- and out-states for the charge conservation on the CFT side, by taking the independent parts of these ambiguities, we checked the correspondence of the deformed disk and annulus amplitudes to the computation on the CFT side.

As a future direction of this work, it is interesting to study refinement of topological B-model on Calabi-Yau threefold. By geometric engineering \cite{Katz:1996fh}, one obtains the correspondence between the topological A-model partition function on some local toric Calabi-Yau threefolds with $A_{N-1}$ singularity and the five dimensional (K-theoretic) version of the Nekrasov partition function with the self-dual condition $\epsilon_1+\epsilon_2=0$ in (\ref{omega}).\footnote{In this correspondence, $\epsilon_1=-\epsilon_2=g_s$ gives us the topological string coupling constant. On the matrix model side, as seen from (\ref{matparam}), the self-dual condition corresponds to $\beta=1$.} This was confirmed in \cite{Iqbal:2003ix, Iqbal:2003zz, Eguchi:2003sj, Eguchi:2003it, Hollowood:2003cv} by the topological vertex \cite{Aganagic:2003db}. Using the refined topological vertex proposed in \cite{Awata:2005fa} and \cite{Iqbal:2007ii}, the extension to general Omega background was also discussed \cite{Awata:2005fa, Iqbal:2007ii, Taki:2007dh, Awata:2008ed}.\footnote{Several world sheet interpretations of refined topological strings are discussed in \cite{Antoniadis:2010iq, Nakayama:2011be, Huang:2011qx}.} In \cite{Gukovtalk}, open string version of geometric engineering was proposed. By this proposal, the ``simple type'' surface operator is embedded to topological open string theory on some local toric Calabi-Yau threefolds with toric brane \cite{Kozcaz:2010af, Dimofte:2010tz, Taki:2010bj, Awata:2010bz}. To consider refinement on the B-model side, matrix model formulation is an important step.\footnote{In \cite{Krefl:2010fm, Huang:2010kf, Krefl:2010jb, Huang:2011qx}, other approaches to the refined B-model are discussed by generalizing the holomorphic anomaly equation \cite{Bershadsky:1993cx, Walcher:2007tp}.} In \cite{Aganagic:2002wv}, for the unrefined A-model on the local toric Calabi-Yau threefold which is the large $N$ dual \cite{Gopakumar:1998ki} of the $U(N)$ Chern-Simons gauge theory on the lens space $S^3/{\IZ}_p$, the Chern-Simons matrix model which has the unitary measure
\begin{equation}
\Delta_q(z):=\prod_{i<j}2\sinh\frac{z_i-z_j}{2}
\end{equation}
instead of the hermitian measure $\Delta(z)$ in (\ref{matrix}) was derived. By the proposal in \cite{Marino:2006hs, Bouchard:2007ys}, it was conjectured that from the Eynard-Orantin recursion \cite{Eynard:2007kz}, one can compute not only the correlation functions with a Wilson loop in the $U(N)$ Chern-Simons gauge theory on $S^3/{\IZ}_p$, but also the unrefined open A-model amplitudes on arbitrary local toric Calabi-Yau threefold with toric brane. In \cite{Brini:2010fc}, for refining this proposal, as a simple guess, it was considered a beta deformed Chern-Simons matrix model:
\begin{equation}
Z_{\beta}:=\frac{1}{N!(2\pi)^N}\int_{\mathbb{R}} \prod_{i=1}^Ndz_i|\Delta_q(z)|^{2\beta}e^{-\frac{\sqrt{\beta}}{g_s}\sum_{i=1}^Nz_i^2}
\label{matrix_b}
\end{equation}
corresponding to the unrefined A-model on the resolved conifold if $\beta=1$. They checked the coincidence between the first deformed disk correlator $W^{(0,1)}_{1}(p)$ obtained from the recursion (\ref{loop_eq}) and a direct perturbative computation. But they also found disagreement with the open refined computation on the A-model side. Therefore we need to reformulate the recursion (\ref{loop_eq}) for obtaining the refined B-model which becomes the mirror dual of the refined A-model, or which gives the K-theoretic version of the Nekrasov partition function in the general Omega background. In \cite{Aganagic:2011sg}, it was proposed a refined Chern-Simons matrix model with the coupling constant dependent measure
\begin{equation}
\Delta_{q,t}(z):=\prod_{m=0}^{\beta-1}\prod_{i<j}(q^{m/2}e^{(z_i-z_j)/2}-q^{-m/2}e^{-(z_i-z_j)/2}),\quad q=e^{g_s},
\end{equation}
which agrees with the open refined A-model computation. It may be important to study the structure of this matrix model for formulating refined B-model.\footnote{See also \cite{Morozov:2012dz} for recent developments of the beta deformation related to this topic.} It is also interesting to approach the problem from the viewpoint of five dimensional version of the AGT correspondence \cite{Awata:2009ur, Awata:2010yy, Awata:2011dc}.

\vspace{2em}
\noindent{\bf Acknowledgements:} I would like to thank Hidetoshi Awata, Hiroyuki Fuji, Hiroaki Kanno and Yasuhiko Yamada for the collaboration \cite{Awata:2010bz} and discussions at the initial stage of this work.

\appendix

\section{Computation on the CFT side}

In this appendix, we summarize the computation on the CFT side corresponding to the $SU(2)$ gauge theories with $N_f=0$, $1$ and $2$ fundamental flavors \cite{Awata:2010bz}.

\subsection{$N_f=0$ case}

The corresponding CFT vacuum state $|\Delta_a, \Lambda\rangle$ which is called the Gaiotto state is defined by \cite{Gaiotto:2009ma},
\begin{equation}
L_1|\Delta_a, \Lambda\rangle=\Lambda^2|\Delta_a, \Lambda\rangle,\quad L_2|\Delta_a, \Lambda\rangle=0,
\end{equation}
where this state is the coherent state of the Virasoro descendant states with the conformal dimension $\Delta_a=\Delta(a)=(b+b^{-1})^2/4-a^2$. The norm of this state coincides with the Nekrasov partition function \cite{Gaiotto:2009ma, Hadasz:2010xp}, and this state is concretely written as \cite{Marshakov:2009gn},
\begin{equation}
|\Delta_a, \Lambda\rangle=\sum_{Y}\Lambda^{2|Y|}Q_{\Delta}^{-1}(1^{|Y|};Y)|\Delta_a,Y \rangle,\quad |\Delta_a,Y \rangle=L_{-\ell}^{n_{\ell}}\cdots L_{-2}^{n_2}L_{-1}^{n_{1}}|\Delta_a \rangle,
\label{G0}
\end{equation}
in terms of the Shapovalov matrix $Q_{\Delta}(Y;Y')$, ($|Y|=|Y'|$) labeled by the Young diagram $Y=(\ell^{n_{\ell}} \ldots 2^{n_2}1^{n_1})$.

We now consider the one point block of the degenerate primary operator $\Phi_{1,2}$ with the conformal dimension $h_{1,2}=-\frac12-\frac{3}{4b^2}$:
\begin{equation}
\Psi^{(1)}(p,a,\Lambda):=
\langle\Delta_-, \Lambda|\Phi_{1,2}(p)|\Delta_+, \Lambda\rangle,\quad \Delta_{\pm}:=\Delta(a \pm \frac{1}{4b}),
\label{oneblock_0}
\end{equation}
where for the charge conservation, we need to shift the momentums of in- and out-states.

By using the null state condition $(b^2L_{-1}^2 + L_{-2}) \Phi_{1,2}(p) =0$, one obtains the differential equation for $\Psi^{(1)}(p,a,\Lambda)=p^{\Delta_--\Delta_+-h_{1,2}}Y^{(1)}(p,a,\Lambda)$ \cite{Awata:2009ur},
\begin{equation}
\Big[\Big(bp\frac{\partial}{\partial p}\Big)^2+2abp\frac{\partial}{\partial p}
+\Lambda^2\Big(p+\frac{1}{p}\Big)+\frac{\Lambda}{4}\frac{\partial}{\partial \Lambda}\Big]Y^{(1)}(p,a,\Lambda)=0.
\label{diff1_0}
\end{equation}
For comparing with the B-model side, after scaling $a \to a/g_s,~ \Lambda \to \Lambda/g_s$, we introduce the parameters $\varepsilon_1$ and $\varepsilon_2$ by (\ref{varep}). And then a series solution $Y^{(1)}(p,a,\Lambda)=1+\sum_{n=1}^{\infty} \Lambda^{2n} Y^{(1)}_n(p,a)$ to the above differential equation is obtained:
\begin{equation}
Y^{(1)}_n(p,a)=\sum_{k=-\infty}^{\infty}A_{n,k}p^k,\quad A_{0,k}=\delta_{0,k},~ A_{n,k}=-\frac{A_{n-1,k-1}+A_{n-1,k+1}}
{\varepsilon_1\left(2ak+\varepsilon_1k^2+\frac{1}{2}n\varepsilon_2\right)}.
\end{equation}
Therefore the free energy is obtained:
\begin{equation}
\sum_{i,j=0}^{\infty}\frac{\varepsilon_1^i\varepsilon_2^j}{\varepsilon_1}{\cal G}^{(1)}_{i,j}(p):=
\log \frac{\Psi^{(1)}(p,a,\Lambda)}{\langle\Delta_a, \Lambda|\Delta_a, \Lambda\rangle}
=\Big(\frac{a}{\varepsilon_1}+\frac12+\frac{3\varepsilon_2}{4\varepsilon_1}\Big)\log p+\log\frac{Y^{(1)}(p,a,\Lambda)}{\langle\Delta_a, \Lambda|\Delta_a, \Lambda\rangle}.
\end{equation}
The free energies ${\cal G}^{(1)}_{i,0}(p)$ obtained by the NS limit $\varepsilon_2 \to 0$ give us the momentum shift invariant solutions in (\ref{oneblock_0}), and the lower order free energies are
\begin{eqnarray}
\label{F00}
{\cal G}^{(1)}_{0,0}(p)&=&a \log p-\frac{p^2-1}{2ap}\Lambda^2-\frac{p^4-1}{16a^3p^2}\Lambda^4-\frac{(p^2-1)(p^4+4p^2+1)}{48a^5p^3}\Lambda^6+{\cal O}(\Lambda^8),\hspace{3em}\\
{\cal G}^{(1)}_{1,0}(p)&=&\frac12\log p+\frac{p^2+1}{4a^2p}\Lambda^2+\frac{(p^2+p+1)(p^2-p+1)}{8a^4p^2}\Lambda^4\nonumber\\
\label{F01}
&&\hspace{8.3em}
+\frac{(p^2+1)(2p^4+p^2+2)}{24a^6p^3}\Lambda^6+{\cal O}(\Lambda^8),\\
\label{F02}
{\cal G}^{(1)}_{2,0}(p)&=&-\frac{p^2-1}{8a^3p}\Lambda^2-\frac{11(p^4-1)}{64a^5p^2}\Lambda^4-\frac{(p^2-1)(10p^4+19p^2+10)}{48a^7p^3}\Lambda^6+{\cal O}(\Lambda^8).
\end{eqnarray}

Next, let us consider the two point block of $\Phi_{1,2}$:
\begin{equation}
\Psi^{(2)}(p_i,a,\Lambda):=
\langle\widetilde{\Delta}_-, \Lambda|\Phi_{1,2}(p_2)\Phi_{1,2}(p_1)|\widetilde{\Delta}_+, \Lambda\rangle,\quad \widetilde{\Delta}_{\pm}:=\Delta(a \pm \frac{1}{2b}).
\end{equation}
As above, one can find the differential equation \cite{Awata:2010bz},
\begin{eqnarray}
&&
\Big[\Big(bp_1\frac{\partial}{\partial p_1}\Big)^2+2abp_1\frac{\partial}{\partial p_1}
+\Lambda^2\Big(p_1+\frac{1}{p_1}\Big)+\frac{\Lambda}{4}\frac{\partial}{\partial \Lambda}\hspace{11.5em}\nonumber\\
&&\hspace{10.75em}
-\frac{p_1+p_2}{2(p_1-p_2)}\Big(p_1\frac{\partial}{\partial p_1}-p_2\frac{\partial}{\partial p_2}\Big)\Big]Y^{(2)}(p_i,a,\Lambda)=0,
\label{diff2_0}
\end{eqnarray}
and the one which is exchanged $p_1$ for $p_2$, where we have defined
\begin{equation}
\Psi^{(2)}(p_i,a,\Lambda)=:p_1^{\Delta_a-\widetilde{\Delta}_+-h_{1,2}}p_2^{\widetilde{\Delta}_--\Delta_a-h_{1,2}}\Big(1-\frac{p_1}{p_2}\Big)^{-\frac{1}{2b^2}}Y^{(2)}(p_i,a,\Lambda)=:f(p_i)Y^{(2)}(p_i,a,\Lambda).
\label{U12}
\end{equation}
By defining (\ref{varep}), a series solution $Y^{(2)}(p_i,a,\Lambda)=1+\sum_{n=1}^{\infty} \Lambda^{2n} Y^{(2)}_n(p_i,a)$ is obtained:
\begin{eqnarray}
&&
Y_n^{(2)}(p_i,a)=\sum_{k_1,k_2=-\infty}^{\infty}A_{n,k_1,k_2}p_1^{k_1}p_2^{k_2},\quad A_{0,k_1,k_2}=\delta_{0,k_1}\delta_{0,k_2},\nonumber\\
&&
A_{n,k_1,k_2}=A_{n,k_2,k_1}=-\frac{A_{n-1,k_1-1,k_2}+A_{n-1,k_1+1,k_2}-A_{n-1,k_1,k_2-1}-A_{n-1,k_1,k_2+1}}
{\varepsilon_1(k_1-k_2)\left(2a+(k_1+k_2)\varepsilon_1\right)},~(k_1\neq k_2),\nonumber\\
&&
A_{n,k,k}=\frac{\varepsilon_1\left(2a(k+1)+\varepsilon_1(k+1)^2+\frac{n+2}{2} \varepsilon_2\right)A_{n,k+1,k-1}-A_{n-1,k+1,k}+A_{n-1,k+2,k-1}}{\varepsilon_1\left(2ak+\varepsilon_1k^2+\frac{n}{2}\varepsilon_2\right)}.\nonumber\\
&&
\end{eqnarray}
The free energy is
\begin{eqnarray}
\sum_{i,j=0}^{\infty}\frac{\varepsilon_1^i\varepsilon_2^j}{\varepsilon_1}{\cal G}^{(2)}_{i,j}(p_1,p_2):&=&
\log \frac{\Psi^{(2)}(p_i,a,\Lambda)}{\langle\Delta_a, \Lambda|\Delta_a, \Lambda\rangle}\hspace{20em}\nonumber\\
&=&
\Big(\frac{a}{\varepsilon_1}+\frac12\Big)\log p_1p_2+\frac{\varepsilon_2}{2\varepsilon_1}\log\frac{p_1^2p_2^2}{p_2-p_1}+\log\frac{Y^{(2)}(p_i,a,\Lambda)}{\langle\Delta_a, \Lambda|\Delta_a, \Lambda\rangle},
\label{free0_2}
\end{eqnarray}
where by comparing the differential equation (\ref{diff2_0}) with (\ref{diff1_0}), we can easily find the relation ${\cal G}^{(2)}_{i,0}(p_1,p_2)={\cal G}^{(1)}_{i,0}(p_1)+{\cal G}^{(1)}_{i,0}(p_2)$ expected from (\ref{NS_rel2}). As discussed in (\ref{shift_inv}), let us take the shift invariant free energies $\widetilde{\cal G}^{(2)}_{i,1}(p_1,p_2)$ of the in- and out-states by neglecting the term $-\frac{\varepsilon_2}{2\epsilon_1}\log(p_1-p_2)$ in (\ref{free0_2}) which is singular at $p_1=p_2$. For example, we get
\begin{eqnarray}
\widetilde{\cal G}^{(2)}_{0,1}(p_1,p_2)&=&
-\frac{(p_1p_2-1)^2}{16a^4p_1p_2}\Lambda^4-\frac{(p_1+p_2)(p_1p_2+1)(p_1^2p_2^2-p_1p_2+1)}{32a^6p_1^2p_2^2}\Lambda^6\hspace{7em}\nonumber\\
&&\hspace{-5em}
-\frac{10(p_1^2+p_2^2)(p_1^4p_2^4+1)+9p_1p_2(p_1^2p_2^2-1)^2
+32p_1^2p_2^2(p_1p_2-1)^2-4p_1^2p_2^2(p_1^2+p_2^2)}{512a^8p_1^3p_2^3}\Lambda^8\nonumber\\
\label{FF00}
&&\hspace{-5em}
+{\cal O}(\Lambda^{10}),\\
\widetilde{\cal G}^{(2)}_{1,1}(p_1,p_2)&=&
\frac{5(p_1^2p_2^2-1)}{32a^5p_1p_2}\Lambda^4+\frac{9(p_1+p_2)(p_1p_2-1)(p_1^2p_2^2+p_1p_2+1)}{64a^7p_1^2p_2^2}\Lambda^6\nonumber\\
&&\hspace{-5em}
+\frac{(p_1^2p_2^2-1)\big(69(p_1^2+p_2^2)(p_1^2p_2^2+1)+61p_1p_2(p_1^2p_2^2+1)+144p_1^2p_2^2\big)}{512a^9p_1^3p_2^3}\Lambda^8+{\cal O}(\Lambda^{10}).
\label{FF01}
\end{eqnarray}

\subsection{$N_f=1$ case}

The Gaiotto state to construct $SU(2)$ gauge theory with one fundamental matter is given by \cite{Gaiotto:2009ma},
\begin{equation}
L_2|\Delta_a, \Lambda,m \rangle=-\Lambda^2|\Delta_a, \Lambda,m \rangle,\quad L_1|\Delta_a, \Lambda\rangle
=-2m \Lambda|\Delta_a, \Lambda,m \rangle,
\end{equation}
where $m$ corresponds to the mass of the fundamental matter. This state is also written as the coherent state of the Virasoro descendant states with the conformal dimension $\Delta_a$ by the Shapovalov matrix as (\ref{G0}) \cite{Marshakov:2009gn}, and the inner product $\langle\Delta_a, \Lambda,m|\Delta_a, \Lambda\rangle$ coincides with the Nekrasov partition function on the gauge theory side \cite{Gaiotto:2009ma, Hadasz:2010xp}.

Here we consider the one point block as
\begin{equation}
\Psi^{(1)}(p,a,m,\Lambda):=\langle\Delta_-, \Lambda,m|\Phi_{1,2}(p)|\Delta_+, 
\Lambda\rangle.
\end{equation}
As in the pure case, $\Psi^{(1)}(p,a,m,\Lambda)=p^{\Delta_--\Delta_+-h_{1,2}}Y^{(1)}(p,a,m,\Lambda)$ satisfies the differential equation
\begin{equation}
\Big[\Big(bp\frac{\partial}{\partial p}\Big)^2+\Big(2ab+\frac16\Big)p
\frac{\partial}{\partial p}-\Lambda^2\Big(p^2-\frac{1}{p}\Big)-2m\Lambda p
+\frac{\Lambda}{3}\frac{\partial}{\partial \Lambda}\Big]Y^{(1)}(p,a,m,\Lambda)=0.
\end{equation}
After scaling $a \to a/g_s$, $\Lambda \rightarrow \Lambda/g_s$, $m \rightarrow m/g_s$, and introducing the parameters (\ref{varep}), one obtains a series solution $Y^{(1)}(p,a,m,\Lambda)=1+\sum_{n=1}^{\infty} \Lambda^n Y^{(1)}_n(p,a,m)$:
\begin{equation}
Y^{(1)}_n(p,a,m)=\sum_{k=-\infty}^{\infty}A_{n,k}p^k,\quad
A_{0,k}=\delta_{0,k},~
A_{n,k}=\frac{A_{n-2,k-2}-A_{n-2,k+1}
+2mA_{n-1,k-1}}{\varepsilon_1\left((2a+\frac16 \varepsilon_2)k + \varepsilon_1k^2+\frac{1}{3}n\epsilon_2\right)}.
\end{equation}
And then we get the free energy
\begin{equation}
\sum_{i,j=0}^{\infty}\frac{\varepsilon_1^i\varepsilon_2^j}{\varepsilon_1}{\cal G}^{(1)}_{i,j}(p):=
\log \frac{\Psi^{(1)}(p,a,m,\Lambda)}{\langle\Delta_a, \Lambda,m|\Delta_a, \Lambda\rangle},
\end{equation}
where the lower order free energies ${\cal G}^{(1)}_{i,0}(p)$ which are the momentum shift invariant are
\begin{eqnarray}
\label{F10}
{\cal G}^{(1)}_{0,0}(p)&=&a\log p+\frac{mp}{a}\Lambda+\frac{(a^2-m^2)p^3+2a^2}{4a^3p}\Lambda^2-\frac{m(a^2-m^2)p^3}
{6a^5}\Lambda^3+{\cal O}(\Lambda^4),\\
{\cal G}^{(1)}_{1,0}(p)&=&\frac12\log p-\frac{mp}{2a^2}\Lambda-\frac{(a^2-2m^2)p^3-a^2}{4a^4p}\Lambda^2+\frac{m\big((6a^2-8m^2)p^3-3a^2\big)}{12a^6}\Lambda^3+{\cal O}(\Lambda^4),\nonumber\\
\label{F11}
&&
\\
\label{F12}
{\cal G}^{(1)}_{2,0}(p)&=&
\frac{mp}{a^3}\Lambda+\frac{(4a^2-11m^2)p^3+2a^2}{16a^5p}\Lambda^2-\frac{5m(5a^2-8m^2)p^3}{24a^7}\Lambda^3+{\cal O}(\Lambda^4).
\end{eqnarray}

We also consider the two point block of $\Phi_{1,2}$:
\begin{equation}
\Psi^{(2)}(p_i,a,m,\Lambda):=
\langle\widetilde{\Delta}_-, \Lambda,m|\Phi_{1,2}(p_2)\Phi_{1,2}(p_1)|\widetilde{\Delta}_+, \Lambda\rangle.
\end{equation}
The differential equations for $\Psi^{(2)}(p_i,a,m,\Lambda)=f(p_i)Y^{(2)}(p_i,a,m,\Lambda)$ are given by \cite{Awata:2010bz},
\begin{eqnarray}
&&
\Big[\Big(bp_1\frac{\partial}{\partial p_1}\Big)^2+\Big(2ab+\frac13\Big)p_1
\frac{\partial}{\partial p_1}-\Lambda^2\Big(p_1^2-\frac{1}{p_1}\Big)-2m\Lambda p_1
+\frac{\Lambda}{3}\frac{\partial}{\partial \Lambda}\hspace{5em}\nonumber\\
&&\hspace{10em}
-\frac{2p_1+p_2}{3(p_1-p_2)}\Big(p_1\frac{\partial}{\partial p_1}-p_2\frac{\partial}{\partial p_2}\Big)\Big]Y^{(2)}(p_i,a,m,\Lambda)=0,
\end{eqnarray}
and the one which is exchanged $p_1$ for $p_2$, where $f(p_i)$ has been defined in (\ref{U12}). By introducing the parameters (\ref{varep}), we obtain a series solution $Y^{(2)}(p_i,a,m,\Lambda)=1+\sum_{n=1}^{\infty} \Lambda^n Y^{(2)}_n(p_i,a,m)$:
\begin{eqnarray}
&&
Y_n^{(2)}(p_i,a,m)=\sum_{k_1,k_2=-\infty}^{\infty}A_{n,k_1,k_2}p_1^{k_1}p_2^{k_2},\quad A_{0,k_1,k_2}=\delta_{0,k_1}\delta_{0,k_2},\nonumber\\
&&
\varepsilon_1(k_1-k_2)\left(2a+(k_1+k_2)\varepsilon_1\right)A_{n,k_1,k_2}\nonumber\\
&&\hspace{6em}
=A_{n-2,k_1-2,k_2}-A_{n-2,k_1+1,k_2}+2mA_{n-1,k_1-1,k_2}-(k_1\leftrightarrow k_2),~(k_1\neq k_2),\nonumber\\
&&
\varepsilon_1\big((2a+\frac13\varepsilon_2)k+\varepsilon_1k^2+\frac{n}{3}\varepsilon_2\big)A_{n,k,k}\nonumber\\
&&\hspace{6em}
=\varepsilon_1\big((2a+\frac{1}{3}\varepsilon_2)(k+1)+\varepsilon_1(k+1)^2+\frac{n+2}{3}\varepsilon_2\big)A_{n,k+1,k-1}\nonumber\\
&&\hspace{8em}
+A_{n-2,k-2,k}-A_{n-2,k-1,k-1}-A_{n-2,k+1,k}+A_{n-2,k+2,k-1}.
\end{eqnarray}
As same as $N_f=0$ case, the free energy
\begin{equation}
\sum_{i,j=0}^{\infty}\frac{\varepsilon_1^i\varepsilon_2^j}{\varepsilon_1}{\cal G}^{(2)}_{i,j}(p_1,p_2):=
\log \frac{\Psi^{(2)}(p_i,a,m,\Lambda)}{\langle\Delta_a, \Lambda|\Delta_a, \Lambda\rangle}.
\end{equation}
is obtained, and we can check the relation ${\cal G}^{(2)}_{i,0}(p_1,p_2)={\cal G}^{(1)}_{i,0}(p_1)+{\cal G}^{(1)}_{i,0}(p_2)$ from the differential equations. The shift invariant free energies $\widetilde{\cal G}^{(2)}_{i,1}(p_1,p_2)$ under the shift (\ref{shift_inv}) are
\begin{eqnarray}
\widetilde{\cal G}^{(2)}_{0,1}(p_1,p_2)&=&
\frac{(a^2-m^2)p_1p_2}{4a^4}\Lambda^2-\frac{m\big((a^2-m^2)p_1p_2(p_1+p_2)+a^2}{4a^6}\Lambda^3\hspace{8em}\nonumber\\
&&\hspace{-4.5em}
-\frac{2(a^2-m^2)(a^2-5m^2)p_1^2p_2^2(p_1^2+p_2^2)+(a^2-m^2)(a^2-9m^2)p_1^3p_2^3+2a^4}{32a^8p_1p_2}\Lambda^4+{\cal O}(\Lambda^5),\nonumber\\
\label{FF10}
&&\\
\widetilde{\cal G}^{(2)}_{1,1}(p_1,p_2)&=&
-\frac{(3a^2-5m^2)p_1p_2}{8a^5}\Lambda^2+\frac{m(7a^2-9m^2)p_1p_2(p_1+p_2)}{8a^7}\Lambda^3\nonumber\\
&&\hspace{-5.5em}
+\frac{(9a^4-70a^2m^2+69m^4)p_1^2p_2^2(p_1^2+p_2^2)+(5a^4-58a^2m^2+61m^4)p_1^3p_2^3-5a^4}{32a^9p_1p_2}\Lambda^4+{\cal O}(\Lambda^5).\nonumber\\
&&
\label{FF11}
\end{eqnarray}

\subsection{$N_f=2$ case}

The irregular CFT corresponding to $SU(2)$ gauge theory with two (anti-)fundamental matters is constructed from the Gaiotto state $|\Delta_a, \Lambda/2,m\rangle$ \cite{Gaiotto:2009ma}. In this case by multiplying $\exp (-\Lambda^2/2\epsilon_1\epsilon_2)$ which is called the $U(1)$ factor by the norm $\langle\Delta_a, \Lambda/2,m_2|\Delta_a, \Lambda/2,m_1\rangle$, this coincides with the Nekrasov partition function \cite{Gaiotto:2009ma, Hadasz:2010xp}.

As same as the previous discussions, the one point block
\begin{equation}
\Psi^{(1)}(p,a,m_i,\Lambda):=\langle\Delta_-, \Lambda,m_2|\Phi_{1,2}(p)
|\Delta_+, \Lambda,m_1\rangle
\end{equation}
satisfies the differential equation,
\begin{equation}
\Big[\Big(bp\frac{\partial}{\partial p}\Big)^2+2abp\frac{\partial}{\partial p}-\Lambda^2\Big(p^2+\frac{1}{p^2}\Big)
-2\Lambda \Big(m_2p+\frac{m_1}{p}\Big)+\frac{\Lambda}{2}\frac{\partial}{\partial \Lambda}\Big]Y^{(1)}(p,a,m_i,\Lambda)=0,
\end{equation}
where we put $\Psi^{(1)}(p,a,m_i,\Lambda)=p^{\Delta_--\Delta_+-h_{1,2}}Y^{(1)}(p,a,m_i,\Lambda)$. After introducing the mass scale $g_s,\varepsilon_1,\varepsilon_2$ as before, one obtains a solution $Y^{(1)}(p,a,m_i,\Lambda)=\sum_{n=1}^{\infty} \Lambda^n Y^{(1)}_n(p,a,m_i)$:
\begin{eqnarray}
&&
Y^{(1)}_n(p,a,m_i)=\sum_{k=-\infty}^{\infty}A_{n,k}p^k,\nonumber\\
&&
A_{0,k}=\delta_{0,k},~ A_{n,k}=\frac{A_{n-2,k-2}+A_{n-2,k+2}+2(m_2A_{n-1,k-1}+m_1A_{n-1,k+1})}
{\varepsilon_1\left(2ak+\varepsilon_1k^2+\frac{1}{2}n\varepsilon_2\right)}.
\end{eqnarray}
The free energy is
\begin{equation}
\sum_{i,j=0}^{\infty}\frac{\varepsilon_1^i\varepsilon_2^j}{\varepsilon_1}{\cal G}^{(1)}_{i,j}(p):=
\log\frac{\Psi^{(1)}(p,a,m_i,\Lambda)}{\langle\Delta_a, \Lambda,m_2|\Delta_a, \Lambda,m_1\rangle},
\end{equation}
where the lower order free energies ${\cal G}^{(1)}_{i,0}(p)$ which are the momentum shift invariant are
\begin{eqnarray}
\label{F20}
{\cal G}^{(1)}_{0,0}(p)&=&a\log p-\frac{m_1-m_2p^2}{ap}\Lambda+\frac{m_1^2-a^2+(a^2-m_2^2)p^4}{4a^3p^2}\Lambda^2+{\cal O}(\Lambda^3),\\
{\cal G}^{(1)}_{1,0}(p)&=&\frac12\log p-\frac{m_1+m_2p^2}{2a^2p}\Lambda-\frac{a^2-2m_1^2-2m_1m_2p^2+(a^2-2m_2^2)p^4}{4a^4p^2}\Lambda^2+{\cal O}(\Lambda^3),\nonumber\\
\label{F21}
&&
\\
\label{F22}
{\cal G}^{(1)}_{2,0}(p)&=&
-\frac{m_1-m_2p^2}{4a^3p}\Lambda+\frac{11m_1^2-4a^2+(4a^2-11m_2^2)p^4}{16a^5p^2}\Lambda^2+{\cal O}(\Lambda^3).
\end{eqnarray}

\section{Functional formula for the Bergman kernel}

The formula (\ref{Bergman}) of the Bergman kernel for genus one Seiberg-Witten curve as
\begin{equation}
y(p)=M(p)\sqrt{\sigma(p)},\quad \sigma(p)=p^4-S_1p^3+S_2p^2-S_3p+S_4=\prod_{i=1}^4(p-q_i),
\label{1SWcurve}
\end{equation}
depends on the branch points of the spectral curve because of the factor
\begin{equation}
G(k)=-\frac16S_2+\frac12(q_1q_2+q_3q_4)-\frac{E(k)}{2K(k)}(q_1-q_3)(q_2-q_4).
\label{BGk}
\end{equation}
In this appendix, by the same discussion to \cite{Bouchard:2008gu, Manabe:2009sf} for local toric del Pezzo surfaces, we prove that this factor can be written as a branch points independent form.

At first, we note the formula
\begin{eqnarray}
&&
K(k)E(k)=\pi^2\Big(\frac{E_2(\tau)}{12}+\omega_1^2e_3\Big),\\
&&
\omega_1:=\frac{2i}{\pi}\frac{K(k)}{\sqrt{(q_1-q_3)(q_2-q_4)}}=\frac{1}{2\pi}\oint_{{\cal A}_1}\frac{dp}{\sqrt{\sigma(p)}},\quad e_3:=\frac{1}{12}\big(S_2-3(q_1q_2+q_3q_4)\big),\nonumber
\end{eqnarray}
where $\tau$ is the elliptic modulus of the curve (\ref{1SWcurve}), and by making use of the prepotential ${\cal F}^{(0,0)}_0$, this is given as
\begin{equation}
\tau=c_1\frac{\partial^2{\cal F}^{(0,0)}_0}{\partial a^2},
\label{ellipt_modulus}
\end{equation}
where $c_1$ is a constant. By this formula, we can rewrite the factor $G(k)$ as \cite{Bouchard:2008gu},
\begin{equation}
G(k)=\frac{E_2(\tau)}{6\omega_1^2}.
\label{Geisen}
\end{equation}
At first, let us note that in Seiberg-Witten theory, the period $\omega_1$ gives us
\begin{equation}
\omega_1=c_2\Lambda\frac{da}{du},
\label{period1}
\end{equation}
where $c_2$ is a constant. In general, the second Eisenstein series $E_2(\tau)$ is written as \cite{Manabe:2009sf},
\begin{equation}
E_2(\tau)=\frac{d}{d\tau}\big(12\log \omega_1+\log \Delta\big),
\label{eisen}
\end{equation}
by the period $\omega_1$, and the discriminant $\Delta=\prod_{i<j}(q_i-q_j)^2=\prod_{i=1}^4\sigma'(q_i)$ of the curve (\ref{1SWcurve}). By (\ref{ellipt_modulus}), the right hand side of (\ref{eisen}) is rewritten as
\begin{equation}
\frac{1}{c_1C_{uuu}}\big(\frac{da}{du}\big)^2\frac{\partial}{\partial u}\big(12\log \omega_1+\log \Delta\big),
\end{equation}
where $C_{uuu}=\frac{\partial^3{\cal F}^{(0,0)}_0}{\partial u^3}$ is the Yukawa coupling. Therefore from (\ref{Geisen}) and (\ref{period1}), we get the functional form by the period of the curve (\ref{1SWcurve}):
\begin{equation}
G(k)=\frac{1}{c\Lambda^2C_{uuu}}\Big(12\log \big(\frac{da}{du}\big)+\log \Delta\Big),\quad c=6c_1c_2^2.
\label{Berg_formula}
\end{equation}

For the $SU(2)$ theories discussed in section 4, the Yukawa couplings are obtained using the discriminants $\Delta$ of the Seiberg-Witten curves (\ref{SW_curve0}), (\ref{SW_curve1}), and (\ref{SW_curve2}) as
\begin{eqnarray}
&&
N_f=0:\quad C_{uuu}=\frac{1}{\Delta\Lambda^4}=\frac{1}{u^2-4\Lambda^4},\hspace{17em}\\
&&
N_f=1:\quad C_{uuu}=\frac{4m^2-3u}{\Delta\Lambda^6}=\frac{4m^2-3u}{4u^2(m^2-u)+4m(8m^2-9u)\Lambda^3-27\Lambda^6},\\
&&
N_f=2:\quad C_{uuu}=\frac{4\big(4m_1^2m_2^2-3u(m_1^2+m_2^2)+2u^2+4m_1m_2\Lambda^2-8\Lambda^4\big)}{\Delta\Lambda^8}.
\end{eqnarray}
The constant $c$ in (\ref{Berg_formula}) can be fixed by estimating the leading behavior of the expansion for the dynamical scale parameter $\Lambda$. For all the cases in section 4, this factor behaves as $G(k)\sim -\frac{1}{6}S_2=-\frac{u}{6\Lambda^2}$, and then the constant is fixed as $c=24$. We can directly check the agreement between (\ref{BGk}) and the formula (\ref{Berg_formula}).

\section{Computation on the B-model side}

In this appendix, we summarize the B-model computation in section 4 and 5.

\subsection{Proof of $W^{(0,1)}_{1,B}(p)=0$ in section 4}

Here, we prove that $W^{(0,1)}_{1,B}(p)$ defined in (\ref{disk0_cor}) vanishes for $N_f=0,1$ and $2$ cases. Let us consider the Seiberg-Witten curve
\begin{equation}
y(p)=M(p)\sqrt{\sigma(p)},\quad \sigma(p)=\prod_{i=1}^4(p-q_i),~M(p)=\frac{\Lambda}{p^2},
\label{SW_curveC}
\end{equation}
where note that for $N_f=1$ and $2$ cases, as noticed in section 4.2 and 4.3 we should take the cycles ${\cal A}_1$ and ${\cal A}_2$ as containing $\infty$ and $0$ respectively. For that proof, we use the fact \cite{Brini:2010fc} that by the M\"obius transformation satisfying $g(g(q))=q$ such as
\begin{eqnarray}
&&
q ~\to~ {\widetilde q}=g(q):=\frac{\alpha q+\beta}{\gamma q-\alpha},\nonumber\\
&&
\alpha:=q_1q_4-q_2q_3,~\beta:=(q_1+q_4)q_2q_3-(q_2+q_3)q_1q_4,~\gamma:=q_1+q_4-q_2-q_3,
\end{eqnarray}
$C^{\scriptsize\mbox{reg}}(q)$ and $C(q)$ in (\ref{dEqL}) are related each other as
\begin{equation}
\Big(\frac{dg(q)}{dq}\Big)LC^{\scriptsize\mbox{reg}}({\widetilde q})
=LC(q)-\frac{\gamma}{\alpha-\gamma q},\quad \Big(\frac{dg(q)}{dq}\Big)LC({\widetilde q})
=LC^{\scriptsize\mbox{reg}}(q)-\frac{\gamma}{\alpha-\gamma q}.
\label{mobius}
\end{equation}
By this M\"obius transformation, the branch points $q_1,q_2,q_3$ and $q_4$ are mapped to $q_4,q_3,q_2$ and $q_1$ respectively, and then the cycle ${\cal A}_1$ is mapped to the cycle ${\cal A}_2$, and vice versa. Therefore for $N_f=1$ and $2$ cases, the cycles ${\cal A}_1$ and ${\cal A}_2$ also contain $-\beta/\alpha$ and $\alpha/\gamma$ respectively. Using the M\"obius transformation (\ref{mobius}), we see that
\begin{equation}
W^{(0,1)}_{1,B}(p)=\frac{Ldp}{8\pi i\sqrt{\sigma(p)}}\oint_{{\cal A}_1}
\Big(\frac{y'(q)}{M(q)}+\frac{y'({\widetilde q})}{M({\widetilde q})}\Big)\frac{dq}{\sqrt{\sigma(q)}},
\label{01Bp}
\end{equation}
where we have used $C^{\scriptsize\mbox{reg}}(q)-C(q)=\frac{1}{\sqrt{\sigma(q)}}$. By
\begin{eqnarray}
\label{IntC1}
&&
\frac{1}{\sqrt{\sigma(q)}}\frac{y'(q)}{M(q)}=-\frac{2}{q}+\frac{1}{2(q-q_1)}+\frac{1}{2(q-q_2)}+\frac{1}{2(q-q_3)}+\frac{1}{2(q-q_4)},\hspace{5em}\\
&&
\frac{1}{\sqrt{\sigma(q)}}\frac{y'({\widetilde q})}{M({\widetilde q})}=\frac{1}{\gamma q-\alpha}\Big[\frac{2(\alpha^2+\beta \gamma)}{\alpha q+\beta}+\frac{(q_1-q_2)(q_1-q_3)}{2(q-q_1)}+\frac{(q_1-q_2)(q_2-q_4)}{2(q-q_2)}\nonumber\\
&&\hspace{15.8em}
+\frac{(q_1-q_3)(q_3-q_4)}{2(q-q_3)}+\frac{(q_2-q_4)(q_3-q_4)}{2(q-q_4)}\Big],
\label{IntC2}
\end{eqnarray}
we see that for $N_f=0$ case, the integrand of (\ref{01Bp}) has the following residues at $q=q_1=\alpha/\gamma=0$ and $q_2$:
\begin{equation}
-\frac32+\frac12~\mbox{for}~(\ref{IntC1}),\quad \frac12+\frac12~\mbox{for}~(\ref{IntC2}),
\end{equation}
and for $N_f=1$ and $2$ cases, it has the following residues at $q=q_1,q_2,\infty$ and $-\beta/\alpha$:
\begin{equation}
\frac12+\frac12+0+0~\mbox{for}~(\ref{IntC1}),\quad \frac12+\frac12+0-2~\mbox{for}~(\ref{IntC2}).
\end{equation}
By summing up these residues, we see that $W^{(0,1)}_{1,B}(p)$ vanishes.

\subsection{Computation of the ``refined'' topological recursion}

In the two-cut case with two sheet as (\ref{1SWcurve}), the constituents of the ``refined'' topological recursion (\ref{loop_eq}) are concretely given by (\ref{dEq}) and (\ref{Bergman}), and especially we need to estimate the complete elliptic integral of the third kind in (\ref{dEq}). As discussed in \cite{Brini:2010fc}, we can treat the ${\cal A}$-period integral in the recursion (\ref{loop_eq}) as the small cut expansion by using the formula:
\begin{eqnarray}
&&
K(k)=\frac{\pi}{2}\sum_{m=0}^{\infty}\Big(\frac{(2m-1)!!}{2^mm!}\Big)^2k^{2m},\\
&&
\Pi(n,k)=\frac{\pi}{2}\sum_{m=0}^{\infty}\frac{\big(\frac12\big)_m^2}{(m!)^2}\bigg[\frac{m!}{\sqrt{1-n}n^m\big(\frac12\big)_m}-\frac{2m}{n}\sum_{j=0}^{m-1}\frac{(1-m)_j}{\big(\frac32\big)_j}\Big(1-\frac{1}{n}\Big)^j\bigg]k^{2m},
\end{eqnarray}
where $(x)_n=\Gamma(x+n)/\Gamma(x)$ is the Pochhammer symbol. From (\ref{dEqL}), because
\begin{equation}
1-n_1=\frac{(q_3-q_2)(q-q_4)}{(q_4-q_2)(q-q_3)},\quad 1-n_4=\frac{(q_3-q_2)(q-q_1)}{(q_3-q_1)(q-q_2)},
\end{equation}
we see that $C^{\scriptsize\mbox{reg}}(q)$ and $C(q)$ have the branch cut only on ${\cal C}_2=[q_3,q_4]$ and ${\cal C}_1=[q_1,q_2]$ respectively. Therefore in the recursion (\ref{loop_eq}), when
\begin{eqnarray}
\widetilde{W}^{(g-1,h+2)}_{\ell}(q,q,p_H)+&&\sum_{k=0}^{g}\sum_{n=0}^{\ell}\sum_{\emptyset=J\subseteq H}\widetilde{W}^{(g-k,|J|+1)}_{\ell-n}(q,p_J)\widetilde{W}^{(k,|H|-|J|+1)}_{n}(q,p_{H \backslash J})\hspace{2em}\nonumber\\
&&\hspace{14.5em}
+dq\frac{d}{dq}\widetilde{W}^{(g,h+1)}_{\ell-1}(q,p_H)
\label{IntegrandK}
\end{eqnarray}
does not have branch cut for $q$, the integrand of (\ref{loop_eq}) also does not have branch cut, and then the ${\cal A}$-period integral can be rewritten as the summation of the branch points $q_i$:
\begin{equation}
\oint_{\cal A}\frac{dq}{2\pi i}\quad\longrightarrow\quad
\sum_{i=1}^4\mathop{\mbox{Res}}_{q=q_i},
\end{equation}
where in the perturbative computation in section 4, for $N_f=1$ and $2$ cases, because the cycle ${\cal A}$ contains $0$ and $\infty$, we also need to take these residues. We see that this is the case of $\ell=$ even number. On the other hand, when (\ref{IntegrandK}) has the branch cuts on ${\cal C}_1$ and ${\cal C}_2$ for $q$, or $\ell$ is odd number, the integrand of (\ref{loop_eq}) also has the branch cuts. In these cases, using the trick as
\begin{eqnarray}
\oint_{{\cal A}_1}\frac{dq}{(q-q_2)^n\sqrt{\sigma(q)}}&=&
\frac{2^n}{(2n-1)!!}\frac{\partial^n}{\partial q_2^n}\oint_{{\cal A}_1}\frac{dq}{\sqrt{\sigma(q)}} \hspace{13em}\nonumber\\
&=&
\frac{2^n}{(2n-1)!!}\frac{\partial^n}{\partial q_2^n}\int_{q_1}^{q_2}\frac{(-2i)dq}{\sqrt{(q-q_1)(q_2-q)(q_3-q)(q_4-q)}},
\end{eqnarray}
we can compute the period.



\begin{thebibliography}{99}
\parskip=-2pt

\bibitem{Alday:2009aq}
  L.~F.~Alday, D.~Gaiotto and Y.~Tachikawa,
  ``Liouville Correlation Functions from Four-dimensional Gauge Theories,''
  Lett.\ Math.\ Phys.\  {\bf 91}, 167 (2010)
  [arXiv:0906.3219 [hep-th]].


\bibitem{Gaiotto:2009we}
  D.~Gaiotto,
  ``N=2 dualities,''
  arXiv:0904.2715 [hep-th].


\bibitem{Nekrasov:2002qd}
  N.~A.~Nekrasov,
  ``Seiberg-Witten Prepotential From Instanton Counting,''
  Adv.\ Theor.\ Math.\ Phys.\  {\bf 7}, 831 (2004)
  [arXiv:hep-th/0206161].


\bibitem{Nekrasov:2010ka}
  N.~Nekrasov and E.~Witten,
  ``The Omega Deformation, Branes, Integrability, and Liouville Theory,''
  JHEP {\bf 1009}, 092 (2010)
  [arXiv:1002.0888 [hep-th]].


\bibitem{Wyllard:2009hg}
  N.~Wyllard,
  ``$A_{N-1}$ conformal Toda field theory correlation functions from conformal
  N=2 SU(N) quiver gauge theories,''
  JHEP {\bf 0911}, 002 (2009)
  [arXiv:0907.2189 [hep-th]].


\bibitem{Gaiotto:2009ma}
  D.~Gaiotto,
  ``Asymptotically free N=2 theories and irregular conformal blocks,''
  arXiv:0908.0307 [hep-th].


\bibitem{Marshakov:2009gn}
  A.~Marshakov, A.~Mironov and A.~Morozov,
  ``On non-conformal limit of the AGT relations,''
  Phys.\ Lett.\  B {\bf 682}, 125 (2009)
  [arXiv:0909.2052 [hep-th]].


\bibitem{Taki:2009zd}
  M.~Taki,
  ``On AGT Conjecture for Pure Super Yang-Mills and W-algebra,''
  JHEP {\bf 1105}, 038 (2011)
  [arXiv:0912.4789 [hep-th]].


\bibitem{Alday:2009fs}
  L.~F.~Alday, D.~Gaiotto, S.~Gukov, Y.~Tachikawa and H.~Verlinde,
   ``Loop and surface operators in N=2 gauge theory and Liouville modular
  geometry,''
  JHEP {\bf 1001}, 113 (2010)
  [arXiv:0909.0945 [hep-th]].


\bibitem{Kozcaz:2010af}
  C.~Kozcaz, S.~Pasquetti and N.~Wyllard,
  ``A \& B model approaches to surface operators and Toda theories,''
  JHEP {\bf 1008}, 042 (2010)
  [arXiv:1004.2025 [hep-th]].


\bibitem{Dimofte:2010tz}
  T.~Dimofte, S.~Gukov and L.~Hollands,
  ``Vortex Counting and Lagrangian 3-manifolds,''
  Lett.\ Math.\ Phys.\  {\bf 98}, 225 (2011)
  [arXiv:1006.0977 [hep-th]].


\bibitem{Dijkgraaf:2009pc}
  R.~Dijkgraaf and C.~Vafa,
  ``Toda Theories, Matrix Models, Topological Strings, and N=2 Gauge Systems,''
  arXiv:0909.2453 [hep-th].


\bibitem{Chekhov:2006rq}
  L.~Chekhov and B.~Eynard,
  ``Matrix eigenvalue model: Feynman graph technique for all genera,''
  JHEP {\bf 0612}, 026 (2006)
  [arXiv:math-ph/0604014].


\bibitem{Chekhov:2010xj}
  L.~Chekhov,
  ``Logarithmic potential beta-ensembles and Feynman graphs,''
  arXiv:1009.5940 [math-ph].


\bibitem{Brini:2010fc}
  A.~Brini, M.~Marino and S.~Stevan,
  ``The uses of the refined matrix model recursion,''
  arXiv:1010.1210 [hep-th].


\bibitem{Eynard:2007kz}
  B.~Eynard and N.~Orantin,
  ``Invariants of algebraic curves and topological expansion,''
  arXiv:math-ph/0702045.


\bibitem{Awata:2010bz}
  H.~Awata, H.~Fuji, H.~Kanno, M.~Manabe and Y.~Yamada,
  ``Localization with a Surface Operator, Irregular Conformal Blocks and Open
  Topological String,''
  arXiv:1008.0574 [hep-th].


\bibitem{Marshakov:2010fx}
  A.~Marshakov, A.~Mironov and A.~Morozov,
  ``On AGT Relations with Surface Operator Insertion and Stationary Limit of
  Beta-Ensembles,''
  J.\ Geom.\ Phys.\  {\bf 61}, 1203 (2011)
  [arXiv:1011.4491 [hep-th]].


\bibitem{Nekrasov:2009rc}
  N.~A.~Nekrasov and S.~L.~Shatashvili,
  ``Quantization of Integrable Systems and Four Dimensional Gauge Theories,''
  arXiv:0908.4052 [hep-th].


\bibitem{Gorsky:1995zq}
  A.~Gorsky, I.~Krichever, A.~Marshakov, A.~Mironov and A.~Morozov,
  ``Integrability and Seiberg-Witten exact solution,''
  Phys.\ Lett.\  B {\bf 355}, 466 (1995)
  [arXiv:hep-th/9505035].


\bibitem{Donagi:1995cf}
  R.~Donagi and E.~Witten,
  ``Supersymmetric Yang-Mills Theory And Integrable Systems,''
  Nucl.\ Phys.\  B {\bf 460}, 299 (1996)
  [arXiv:hep-th/9510101].


\bibitem{Mironov:2009uv}
  A.~Mironov and A.~Morozov,
  ``Nekrasov Functions and Exact Bohr-Sommerfeld Integrals,''
  JHEP {\bf 1004}, 040 (2010)
  [arXiv:0910.5670 [hep-th]].


\bibitem{Mironov:2009dv}
  A.~Mironov and A.~Morozov,
  ``Nekrasov Functions from Exact BS Periods: the Case of SU(N),''
  J.\ Phys.\ A  {\bf 43}, 195401 (2010)
  [arXiv:0911.2396 [hep-th]].


\bibitem{Mironov:2009ib}
  A.~Mironov, A.~Morozov and S.~Shakirov,
  ``Matrix Model Conjecture for Exact BS Periods and Nekrasov Functions,''
  JHEP {\bf 1002}, 030 (2010)
  [arXiv:0911.5721 [hep-th]].


\bibitem{Popolitov:2010bz}
  A.~Popolitov,
  ``On relation between Nekrasov functions and BS periods in pure SU(N) case,''
  arXiv:1001.1407 [hep-th].


\bibitem{Teschner:2010je}
  J.~Teschner,
  ``Quantization of the Hitchin moduli spaces, Liouville theory, and the
  geometric Langlands correspondence I,''
  arXiv:1005.2846 [hep-th].


\bibitem{He:2010xa}
  W.~He and Y.~G.~Miao,
  ``Magnetic expansion of Nekrasov theory: the SU(2) pure gauge theory,''
  Phys.\ Rev.\  D {\bf 82}, 025020 (2010)
  [arXiv:1006.1214 [hep-th]].


\bibitem{Maruyoshi:2010iu}
  K.~Maruyoshi and M.~Taki,
  ``Deformed Prepotential, Quantum Integrable System and Liouville Field
  Theory,''
  Nucl.\ Phys.\  B {\bf 841}, 388 (2010)
  [arXiv:1006.4505 [hep-th]].


\bibitem{Poghossian:2010pn}
  R.~Poghossian,
  ``Deforming SW curve,''
  JHEP {\bf 1104}, 033 (2011)
  [arXiv:1006.4822 [hep-th]].


\bibitem{Tai:2010ps}
  T.~S.~Tai,
  ``Uniformization, Calogero-Moser/Heun duality and Sutherland/bubbling
  pants,''
  JHEP {\bf 1010}, 107 (2010)
  [arXiv:1008.4332 [hep-th]].


\bibitem{Piatek:2011tp}
  M.~Piatek,
  ``Classical conformal blocks from TBA for the elliptic Calogero-Moser
  system,''
  JHEP {\bf 1106}, 050 (2011)
  [arXiv:1102.5403 [hep-th]].


\bibitem{Nekrasov:2011bc}
  N.~Nekrasov, A.~Rosly and S.~Shatashvili,
  ``Darboux coordinates, Yang-Yang functional, and gauge theory,''
  Nucl.\ Phys.\ Proc.\ Suppl.\  {\bf 216}, 69 (2011)
  [arXiv:1103.3919 [hep-th]].


\bibitem{Fucito:2011pn}
  F.~Fucito, J.~F.~Morales, D.~R.~Pacifici and R.~Poghossian,
  ``Gauge theories on Omega-backgrounds from non commutative Seiberg-Witten
  curves,''
  JHEP {\bf 1105}, 098 (2011)
  [arXiv:1103.4495 [hep-th]].


\bibitem{Zenkevich:2011zx}
  Y.~Zenkevich,
  ``Nekrasov prepotential with fundamental matter from the quantum spin
  chain,''
  Phys.\ Lett.\  B {\bf 701}, 630 (2011)
  [arXiv:1103.4843 [math-ph]].


\bibitem{Dorey:2011pa}
  N.~Dorey, S.~Lee and T.~J.~Hollowood,
  ``Quantization of Integrable Systems and a 2d/4d Duality,''
  JHEP {\bf 1110}, 077 (2011)
  [arXiv:1103.5726 [hep-th]].


\bibitem{Chen:2011sj}
  H.~Y.~Chen, N.~Dorey, T.~J.~Hollowood and S.~Lee,
  ``A New 2d/4d Duality via Integrability,''
  JHEP {\bf 1109}, 040 (2011)
  [arXiv:1104.3021 [hep-th]].


\bibitem{Bonelli:2011na}
  G.~Bonelli, K.~Maruyoshi and A.~Tanzini,
  ``Quantum Hitchin Systems via beta-deformed Matrix Models,''
  arXiv:1104.4016 [hep-th].


\bibitem{Aganagic:2011mi}
  M.~Aganagic, M.~C.~N.~Cheng, R.~Dijkgraaf, D.~Krefl and C.~Vafa,
  ``Quantum Geometry of Refined Topological Strings,''
  arXiv:1105.0630 [hep-th].


\bibitem{Huang:2011qx}
  M.~x.~Huang, A.~K.~Kashani-Poor and A.~Klemm,
  ``The Omega deformed B-model for rigid N=2 theories,''
  arXiv:1109.5728 [hep-th].


\bibitem{Migdal:1984gj}
  A.~A.~Migdal,
  ``Loop Equations And 1/N Expansion,''
  Phys.\ Rept.\  {\bf 102}, 199 (1983).


\bibitem{Ambjorn:1992gw}
  J.~Ambjorn, L.~Chekhov, C.~F.~Kristjansen and Yu.~Makeenko,
  ``Matrix model calculations beyond the spherical limit,''
  Nucl.\ Phys.\  B {\bf 404}, 127 (1993)
  [Erratum-ibid.\  B {\bf 449}, 681 (1995)]
  [arXiv:hep-th/9302014].


\bibitem{Akemann:1996zr}
  G.~Akemann,
  ``Higher genus correlators for the Hermitian matrix model with multiple
  cuts,''
  Nucl.\ Phys.\  B {\bf 482}, 403 (1996)
  [arXiv:hep-th/9606004].


\bibitem{Eynard:2004mh}
  B.~Eynard,
  ``Topological expansion for the 1-hermitian matrix model correlation
  functions,''
  JHEP {\bf 0411}, 031 (2004)
  [arXiv:hep-th/0407261].


\bibitem{Alexandrov:2003pj}
  A.~S.~Alexandrov, A.~Mironov and A.~Morozov,
  ``Partition functions of matrix models as the first special functions of
  string theory. I: Finite size Hermitean 1-matrix model,''
  Int.\ J.\ Mod.\ Phys.\  A {\bf 19}, 4127 (2004)
  [Teor.\ Mat.\ Fiz.\  {\bf 142}, 419 (2005)]
  [arXiv:hep-th/0310113].


\bibitem{Bouchard:2007ys}
  V.~Bouchard, A.~Klemm, M.~Marino and S.~Pasquetti,
  ``Remodeling the B-model,''
  Commun.\ Math.\ Phys.\  {\bf 287}, 117 (2009)
  [arXiv:0709.1453 [hep-th]].


\bibitem{Kostov:1999xi}
  I.~K.~Kostov,
  ``Conformal field theory techniques in random matrix models,''
  arXiv:hep-th/9907060.


\bibitem{Aganagic:2003qj}
  M.~Aganagic, R.~Dijkgraaf, A.~Klemm, M.~Marino and C.~Vafa,
  ``Topological strings and integrable hierarchies,''
  Commun.\ Math.\ Phys.\  {\bf 261}, 451 (2006)
  [arXiv:hep-th/0312085].


\bibitem{Eguchi:2009gf}
  T.~Eguchi and K.~Maruyoshi,
  ``Penner Type Matrix Model and Seiberg-Witten Theory,''
  JHEP {\bf 1002}, 022 (2010)
  [arXiv:0911.4797 [hep-th]].


\bibitem{Fujita:2009gf}
  M.~Fujita, Y.~Hatsuda and T.~S.~Tai,
  ``Genus-one correction to asymptotically free Seiberg-Witten prepotential
  from Dijkgraaf-Vafa matrix model,''
  JHEP {\bf 1003}, 046 (2010)
  [arXiv:0912.2988 [hep-th]].


\bibitem{Katz:1996fh}
  S.~H.~Katz, A.~Klemm and C.~Vafa,
  ``Geometric engineering of quantum field theories,''
  Nucl.\ Phys.\  B {\bf 497}, 173 (1997)
  [arXiv:hep-th/9609239].


\bibitem{Iqbal:2003ix}
  A.~Iqbal and A.~K.~Kashani-Poor,
  ``Instanton counting and Chern-Simons theory,''
  Adv.\ Theor.\ Math.\ Phys.\  {\bf 7}, 457 (2004)
  [arXiv:hep-th/0212279].


\bibitem{Iqbal:2003zz}
  A.~Iqbal and A.~K.~Kashani-Poor,
  ``SU(N) geometries and topological string amplitudes,''
  Adv.\ Theor.\ Math.\ Phys.\  {\bf 10}, 1 (2006)
  [arXiv:hep-th/0306032].


\bibitem{Eguchi:2003sj}
  T.~Eguchi and H.~Kanno,
  ``Topological strings and Nekrasov's formulas,''
  JHEP {\bf 0312}, 006 (2003)
  [arXiv:hep-th/0310235].


\bibitem{Eguchi:2003it}
  T.~Eguchi and H.~Kanno,
   ``Geometric transitions, Chern-Simons gauge theory and Veneziano type
  amplitudes,''
  Phys.\ Lett.\  B {\bf 585}, 163 (2004)
  [arXiv:hep-th/0312234].


\bibitem{Hollowood:2003cv}
  T.~J.~Hollowood, A.~Iqbal and C.~Vafa,
  ``Matrix Models, Geometric Engineering and Elliptic Genera,''
  JHEP {\bf 0803}, 069 (2008)
  [arXiv:hep-th/0310272].


\bibitem{Aganagic:2003db}
  M.~Aganagic, A.~Klemm, M.~Marino and C.~Vafa,
  ``The topological vertex,''
  Commun.\ Math.\ Phys.\  {\bf 254}, 425 (2005)
  [arXiv:hep-th/0305132].


\bibitem{Awata:2005fa}
  H.~Awata and H.~Kanno,
   ``Instanton counting, Macdonald functions and the moduli space of
  D-branes,''
  JHEP {\bf 0505}, 039 (2005)
  [arXiv:hep-th/0502061].


\bibitem{Iqbal:2007ii}
  A.~Iqbal, C.~Kozcaz and C.~Vafa,
  ``The refined topological vertex,''
  JHEP {\bf 0910}, 069 (2009)
  [arXiv:hep-th/0701156].


\bibitem{Taki:2007dh}
  M.~Taki,
  ``Refined Topological Vertex and Instanton Counting,''
  JHEP {\bf 0803}, 048 (2008)
  [arXiv:0710.1776 [hep-th]].


\bibitem{Awata:2008ed}
  H.~Awata and H.~Kanno,
  ``Refined BPS state counting from Nekrasov's formula and Macdonald
  functions,''
  Int.\ J.\ Mod.\ Phys.\  A {\bf 24}, 2253 (2009)
  [arXiv:0805.0191 [hep-th]].


\bibitem{Antoniadis:2010iq} 
  I.~Antoniadis, S.~Hohenegger, K.~S.~Narain and T.~R.~Taylor,
  ``Deformed Topological Partition Function and Nekrasov Backgrounds,''
  Nucl.\ Phys.\ B {\bf 838}, 253 (2010)
  [arXiv:1003.2832 [hep-th]].


\bibitem{Nakayama:2011be}
  Y.~Nakayama and H.~Ooguri,
  ``Comments on Worldsheet Description of the Omega Background,''
  Nucl.\ Phys.\  B {\bf 856}, 342 (2012)
  [arXiv:1106.5503 [hep-th]].


\bibitem{Gukovtalk}
  S.~Gukov,
   ``Surface operators in N=2 gauge theories and duality,''
   talk in the ASC Workshop on interfaces and wall-crossing, Munich, December, 2009.


\bibitem{Taki:2010bj}
  M.~Taki,
  ``Surface Operator, Bubbling Calabi-Yau and AGT Relation,''
  JHEP {\bf 1107}, 047 (2011)
  [arXiv:1007.2524 [hep-th]].


\bibitem{Krefl:2010fm}
  D.~Krefl and J.~Walcher,
  ``Extended Holomorphic Anomaly in Gauge Theory,''
  Lett.\ Math.\ Phys.\  {\bf 95}, 67 (2011)
  [arXiv:1007.0263 [hep-th]].


\bibitem{Huang:2010kf}
  M.~x.~Huang and A.~Klemm,
  ``Direct integration for general Omega backgrounds,''
  arXiv:1009.1126 [hep-th].


\bibitem{Krefl:2010jb}
  D.~Krefl and J.~Walcher,
  ``Shift versus Extension in Refined Partition Functions,''
  arXiv:1010.2635 [hep-th].


\bibitem{Bershadsky:1993cx}
  M.~Bershadsky, S.~Cecotti, H.~Ooguri and C.~Vafa,
  ``Kodaira-Spencer theory of gravity and exact results for quantum string
  amplitudes,''
  Commun.\ Math.\ Phys.\  {\bf 165}, 311 (1994)
  [arXiv:hep-th/9309140].


\bibitem{Walcher:2007tp}
  J.~Walcher,
  ``Extended Holomorphic Anomaly and Loop Amplitudes in Open Topological
  String,''
  Nucl.\ Phys.\  B {\bf 817}, 167 (2009)
  [arXiv:0705.4098 [hep-th]].


\bibitem{Aganagic:2002wv}
  M.~Aganagic, A.~Klemm, M.~Marino and C.~Vafa,
  ``Matrix model as a mirror of Chern-Simons theory,''
  JHEP {\bf 0402}, 010 (2004)
  [arXiv:hep-th/0211098].


\bibitem{Gopakumar:1998ki}
  R.~Gopakumar and C.~Vafa,
  ``On the gauge theory/geometry correspondence,''
  Adv.\ Theor.\ Math.\ Phys.\  {\bf 3}, 1415 (1999)
  [arXiv:hep-th/9811131].


\bibitem{Marino:2006hs}
  M.~Marino,
  ``Open string amplitudes and large order behavior in topological string
  theory,''
  JHEP {\bf 0803}, 060 (2008)
  [arXiv:hep-th/0612127].


\bibitem{Aganagic:2011sg}
  M.~Aganagic and S.~Shakirov,
  ``Knot Homology from Refined Chern-Simons Theory,''
  arXiv:1105.5117 [hep-th].


\bibitem{Morozov:2012dz}
  A.~Morozov,
  ``Challenges of beta-deformation,''
  arXiv:1201.4595 [hep-th].


\bibitem{Awata:2009ur}
  H.~Awata and Y.~Yamada,
  ``Five-dimensional AGT Conjecture and the Deformed Virasoro Algebra,''
  JHEP {\bf 1001}, 125 (2010)
  [arXiv:0910.4431 [hep-th]].


\bibitem{Awata:2010yy}
  H.~Awata and Y.~Yamada,
  ``Five-dimensional AGT Relation and the Deformed beta-ensemble,''
  Prog.\ Theor.\ Phys.\  {\bf 124}, 227 (2010)
  [arXiv:1004.5122 [hep-th]].


\bibitem{Awata:2011dc}
  H.~Awata, B.~Feigin, A.~Hoshino, M.~Kanai, J.~Shiraishi and S.~Yanagida,
  ``Notes on Ding-Iohara algebra and AGT conjecture,''
  arXiv:1106.4088 [math-ph].


\bibitem{Hadasz:2010xp}
  L.~Hadasz, Z.~Jaskolski and P.~Suchanek,
  ``Proving the AGT relation for $N_f = 0,1,2$ antifundamentals,''
  JHEP {\bf 1006}, 046 (2010)
  [arXiv:1004.1841 [hep-th]].


\bibitem{Bouchard:2008gu}
  V.~Bouchard, A.~Klemm, M.~Marino and S.~Pasquetti,
  ``Topological open strings on orbifolds,''
  Commun.\ Math.\ Phys.\  {\bf 296}, 589 (2010)
  [arXiv:0807.0597 [hep-th]].


\bibitem{Manabe:2009sf}
  M.~Manabe,
  ``Topological open string amplitudes on local toric del Pezzo surfaces via
  remodeling the B-model,''
  Nucl.\ Phys.\  B {\bf 819}, 35 (2009)
  [arXiv:0903.2092 [hep-th]].


\end{thebibliography}
\end{document}